\documentclass[english, 12pt]{article}

\RequirePackage[OT1]{fontenc}
\RequirePackage{amsthm,amsmath, amssymb}
\RequirePackage[round, longnamesfirst]{natbib}
\RequirePackage[colorlinks, linkcolor=blue, anchorcolor=blue, citecolor=blue,
filecolor=blue, urlcolor=blue]{hyperref}
\usepackage{bbm}
\usepackage{subfigure}
\usepackage{graphicx}
\usepackage[english]{babel}
\usepackage{setspace}
\usepackage[margin=1.2in]{geometry}

\title{Noncrossing structured additive multiple-output Bayesian quantile regression models}
\author{Bruno Santos$^\star$ and Thomas Kneib$^\ast$}
\date{Georg-August-Universit\"at G\"ottingen \\ \medskip
\small $^\star$ bruno.santos@uni-goettingen.de; \\ $^\ast$ tkneib@uni-goettingen.de}

\begin{document}
\maketitle

\onehalfspacing

\begin{abstract}
Quantile regression models are a powerful tool for studying different points of the conditional distribution of univariate response variables. Their multivariate counterpart extension though is not straightforward, starting with the definition of multivariate quantiles. We propose here a flexible Bayesian quantile regression model when the response variable is multivariate, where we are able to define a structured additive framework for all predictor variables. We build on previous ideas considering a directional approach to define the quantiles of a response variable with multiple-outputs and we define noncrossing quantiles in every directional quantile model. We define a Markov Chain Monte Carlo (MCMC) procedure for model estimation, where the noncrossing property is obtained considering a Gaussian process design to model the correlation between several quantile regression models. We illustrate the results of these models using two data sets: one on dimensions of inequality in the population, such as income and health; the second on scores of students in the Brazilian High School National Exam, considering three dimensions for the response variable.
\end{abstract}

\newpage

\section{Introduction}

Quantiles can be easily defined for univariate probability distributions, providing meaningful information about location and also about dispersion of the data. Considering its representation in a regression framework, one can say that quantile regression models are a great mechanism to describe how a set of predictors might influence a response variable differently depending on the quantile of interest. Take the linear quantile regression, for example, as we assume the following for the conditional quantile of $Y$
\[
 Q_Y(\tau|\boldsymbol X = \boldsymbol x) =  \eta_\tau  = \boldsymbol x^{'} {\boldsymbol \beta_\tau} ,
\]
where $\boldsymbol x$ is a vector of predictor variables and $\boldsymbol \beta_\tau$ is the vector of quantile regression parameters, both of same dimension size. This idea was fully systematized by \citet{koenker:78} and since its inception has been applied in several areas, from Economy to Ecology \citep[see, for instance,][]{koenker:05, yu:03}. Its main use has been for continuous response variables, though there are proposals in the literature for count data \citep{machado2005} and binary response variables \citep{ji:12} as well. Estimation is carried out with linear programming algorithms and inferential procedures, such as confidence intervals, can be achieved via asymptotic results or bootstrap, for instance. The Bayesian approach was first proposed by \citet{yu:01}, where an asymmetric Laplace distribution is proposed for the likelihood. A more efficient representation for this distribution was given by \citet{kozumi:11}, which we will be considered here as well for our estimation purposes. Later, \citet{sriram:13} proved posterior consistency for these models, when considering an asymmetric Laplace distribution, even when the true data generating process is another probability distribution. 

Despite the great attention received for quantile regression models over the years, extensions considering multivariate response variables are more complicated. Given the advantages one has when using quantile regression models, such as the lack of a parametric probability assumption for the response variable, a similar approach for a multiple-output response variable is certainly appealing. Likewise the interest would be identical, for instance, trying to explore the multivariate conditional probability distribution, but taking into account the possible related effects of predictors not on average, but rather in different parts of this distribution, namely different quantiles.

A first challenge though presents itself in the definition of ordering mechanisms for more than one dimension, in the same sense as quantiles. While one can easily define such measures for univariate probability distributions, a definition for similar quantities can be done in different ways for multivariate data. \citet{serfling2002} gives a good introduction in the topic, describing some properties that might be important to these type of measures. One interesting definition relates to the idea of depth in the data and it can be traced back to \citet{tukey1975}. In the univariate case one can characterize the depth of a point $y \in \mathbb{R}$ as $\min(F(y), 1-F(y))$, where $F$ is the cumulative distribution function of Y. Regarding multivariate depth, \citet{mosler2013} summarizes several options for these functions based either on distances, such as Mahalanobis depth and Oja depth, or weighted distances or also on halfspaces. This latter can be connected to the multiple-output quantile regression framework which is the interest of this paper. This linkage between these two ideas was established by \citet{hallin:10}, although a similar effort was pursued by \citet{kong2012} with a related definition for multivariate quantiles. We focus here on the former definition and will not give more details on the latter. For more information on depth functions and multivariate quantiles, we also refer to \citet{chernozhukov2017} and  \citet{carlier2016, carlier2017} for a more recent discussion on the topic.

The same way as one is interested in taking into consideration the multivariate nature of the response variable, one should also acknowledge more flexible structures for the predictor variables. A possible way of achieving this goal is considering structured additive models in the context of quantile regression. In this situation, the predictor $\eta_{\tau}$ for the $\tau$th conditional quantile can be expressed as
\begin{equation*}
 \eta_{\tau} = f_{1\tau}(z_{1}) + \cdots + f_{q\tau}(z_{q}) + \boldsymbol x^{'} \boldsymbol \beta_\tau,
\end{equation*}
where $f_{j\tau}$, $j = 1,\ldots,q$, are nonlinear functions related to covariates $z_1, \ldots, z_q$ and $\boldsymbol x^{'} \boldsymbol \beta_\tau$ is the typical linear quantile regression part, using the covariates vector $\boldsymbol x$. These nonlinear functions might represent time trends for continuous variables, varying coefficient for terms cluster data or even spatial effects. If we drop the index $\tau$ to ease the notation, we can write each $f_j(.)$ as a linear combination of basis such as 
\begin{equation} \label{basisDefinition}
 f_j({z_j}) = \sum_{k=1}^K \gamma_{jk} B_{jk}({z_j}),
\end{equation}
where ${\boldsymbol \gamma_j} = (\gamma_{j1}, \ldots, \gamma_{jK})$ is the vector of unknown coefficients and the $B_{jk}$ are known basis functions. This type of scheme was considered for studying copula models by \citet{klein2016}, for instance. Furthermore, \citet{Lang2014} used this design in the analysis of multilevel hierarchical models. \citet{waldmann2013} considered this approach for Bayesian quantile regression and we use their results in this directional quantile method for multiple-output response variables.

Moreover, when estimating these conditional quantiles separately often there is a problem of crossing quantiles, which violates a basic probability law, that defines that quantiles are nondecreasing functions with respect to $\tau$ for any given covariate. One possible solution for this problem is to estimate these models jointly, defining constraints that guarantee the monotonicity assumption, as it was done by \citet{tokdar:11} for one explanatory variable and by \citet{yang2017} for the multivariate case, in a Bayesian setting.  Similar approaches to estimating simultaneously several quantile regression models were proposed by \citet{bondell2010} and  \citet{reich:10}. The former used a weight function to write a constrained minimization problem and the latter considered Bernstein basis polynomials to define the quantile process.  \citet{reich:10} specified this approach in the light of spatial data problems, but its likelihood evaluation might be impractical for datasets with a larger number of observations. Moreover, \citet{Rodrigues2019} define a new method for joint quantile regression modeling based on quantile pyramids, where the quantile parameters are ensured to respect a noncrossing constraint. \citet{Rodrigues2019b} extended this idea of pyramid quantile regression in the context of a spline regression setting. \citet{he1997} proposed restricted regression quantiles imposing constraints in the space of solutions for the conditional quantiles, attaining the noncrossing property as the result. In a frequentist scenario, \citet{chernozhukov2009} proposes  reordering of the estimated quantiles to obtain monotonic results. \citet{Rodrigues2017} proposed a rearrangement of the estimated Bayesian quantile regression coefficients via a Gaussian process algorithm, which connects the different quantile through its correlation function. This last approach is the one we consider for our method in order to produce noncrossing quantiles in this multiple-output context. 

In this article, we consider a Bayesian framework for the directional method proposed by \citet{hallin:10}, where we are able to use prior information whenever this is available. Furthermore, we add the possibility of using of nonlinear functions to explain the effect of covariates in the response variable, in the form of structured additive predictors. Finally, we are concerned with the problem of crossing quantiles, to which we adapt solutions in the literature for the univariate case to this multiple-output setting.

This rest of the article is organized in the following way. Section~\ref{multOut} defines the Tukey depth concept and its connection to the important results of the multiple-output quantile regression model defined by \citet{hallin:10}. Following, Section~\ref{STADNCQR} introduces the extension to a Bayesian scheme, which was proposed by \citet{guggisberg2017}, and presents our contributions to the literature regarding the necessary adjustments to obtain noncrossing conditional quantiles. We discuss the methods proposed in this paper with an application to two real datasets in Section~\ref{application}. We finish with our last considerations in Section~\ref{finalRemarks}.

\section{Multiple-output Bayesian quantile regression}
\label{multOut}

Considering a response variable defined as ${\boldsymbol Y} \in \mathbb{R}^k$, then a first challenge becomes how one can define an ordering measure for this variable. This definition is important if we are interested in studying how changes in explanatory variables might affect different locations of the response variable, which is one of the goals of quantile regression models, for instance. In fact, this ordering can be achieved in different ways, as discussed by \citet{serfling2002} considering quantile functions. We focus here on the definitions of depth for multivariate data. We refer to \citet{mosler2013} and \citet{chernozhukov2017}, and references therein, for a review on properties and definitions of different depth functions. We concentrate on the Tukey depth, which is connected to the quantile regression approach for the multivariate data we pursue here. One can define the Tukey depth, also known as halfspace depth, of an observation $\boldsymbol z \in \mathbb{R}^k$ with respect to some probability distribution $P$ as 
\begin{equation*}
 HD(\boldsymbol z , P) := \inf\{ P(H): H \mbox{ is a closed halfspace containing } \boldsymbol z \},
\end{equation*}
where halfspace is one of the parts when the space is divided in two parts by a hyperplane. This definition allows one to describe a region of points with depth at least $\tau$, $D(\tau)$, as 
\begin{equation*}
 D(\tau) := \{\boldsymbol z \in \mathbb{R}^k : HD(\boldsymbol z, P) \geq \tau \}.
\end{equation*}

For discrete or continuous random variables with compact support, $D(\tau)$ determines the probability distribution of $Y$ \citep{struyf1999}. In particular, these regions provide a good feature to use in a regression setting. For instance, one can be interested in relating the measure $P$ to different values of the predictor variables and then checking how these depth regions $D(\tau)$ vary. This certainly can be a good indication on how the predictor variables affect the conditional distribution of this multiple-output response variable. But though we are able to define this depth regions for different values of $k$, visualization of these variations might be an issue for $k > 3$. 

Regarding the multiple-output quantile regression approach proposed by \citet{hallin:10}, we need to delineate a few terms. Let a directional index be ${\boldsymbol \tau} \in \mathcal B^k := \{ {\boldsymbol v} \in \mathbb{R}^k: 0 < || {\boldsymbol v} ||_2 < 1 \}$, which is a collection of vectors encompassed in the unit ball of $\mathbb{R}^k$. This directional index can be split into two parts, ${\boldsymbol \tau} = \tau {\boldsymbol u}$, where ${\boldsymbol u} \in \mathcal{S}^{k-1} :=  \{\boldsymbol z \in \mathbb{R}^k : ||\boldsymbol z|| = 1 \}$, represents the direction and $\tau \in (0,1)$ constitutes the 
magnitude. Now let $\boldsymbol \Gamma_u$ be an arbitrary $k \times (k-1)$ matrix of unit vectors, where 
$(\boldsymbol u \, \vdots \, \boldsymbol \Gamma_u)$ establishes an orthonormal basis of $\mathbb{R}^k$. Lastly, let $\rho_\tau(u)$ be the usual check loss function often used in quantile regression analysis, i.e., $\rho_\tau(u) = u (\tau - \mathbb{I}(u < 0))$, where $\mathbb{I}(.)$ is the indicator function. Now we can restate Definition 2.1 by \citet{hallin:10} as the following

\medskip

DEFINITION 2.1. The $\boldsymbol \tau$th quantile of $\boldsymbol Y$ is any element of the collection $\Lambda_\tau$ of hyperplanes $\lambda_\tau := \{ \boldsymbol y \in \mathbb{R}^k : \boldsymbol u^{'} \boldsymbol y = \hat{\boldsymbol b}_\tau \boldsymbol \Gamma^{'}_u \boldsymbol y + \hat{a}_\tau \}$ such that $(\hat{a}_\tau, \hat{\boldsymbol b}_\tau)$ are the solutions of the minimization problem 
\begin{equation} \label{minProb}
  \min_{(a_\tau, \boldsymbol b_\tau) \in \mathbb{R}^k} E[\rho_\tau(\boldsymbol u^{'} \boldsymbol y - \boldsymbol b_\tau \boldsymbol \Gamma^{'}_u \boldsymbol y - a_\tau )].
\end{equation}

These directional quantiles are related to the usual quantile regression estimator of \citet{koenker:78} given the minimization problem in \eqref{minProb}. In fact, Definition 2.1 does not mention predictor variables, but its addition is trivial in the sense that will create new levels of the collection of hyperplanes $\Lambda_\tau$, which will be conditional on the values of $\boldsymbol X$, i.e. 
\begin{equation*} \label{lambda_hyperplane}
\lambda_\tau(\boldsymbol X) = \{ \boldsymbol u^{'} \boldsymbol y = \hat{\boldsymbol b}_\tau \boldsymbol \Gamma^{'}_u \boldsymbol y + 
\boldsymbol x^{'} \hat{\boldsymbol \beta}_\tau  + \hat{a}_\tau \},
\end{equation*}
where $\boldsymbol x$ is vector of dimension $p$ with the values of the explanatory variables and $\hat{\boldsymbol \beta}_\tau$ is the quantile regression estimate in the minimization problem similar as in \eqref{minProb}, the coefficient related to $\boldsymbol X$.

We also can give the empirical version of this approach. For the sake of brevity in the notation consider the following terms,
$ Y_u := {\boldsymbol u}^{'} {\boldsymbol Y}$, ${\boldsymbol Y}^{\perp} := {\boldsymbol \Gamma_u}^{'} \boldsymbol Y$. Taking a sample of $n$ observations of $({\boldsymbol Y_i, \boldsymbol X_i})$ one can find the directional multivariate quantile regression parameters as the solution of the minimization problem
\begin{equation} \label{minization}
 \min_{\alpha_\tau, \beta_{\tau Y}, \beta_{\tau X}} \frac{1}{n} \sum_{i=1}^n \rho_\tau (Y_{iu} - {\boldsymbol 
 \beta}_{\tau Y}^{'} {\boldsymbol Y_{iu}^{\perp}} -  {\boldsymbol \beta_{\tau X}^{'} \boldsymbol X_i } - \alpha_\tau),
\end{equation}
where $\boldsymbol X_i$ is the vector of predictor variables, $\alpha_\tau$ is an intercept term, $\boldsymbol \beta_{\tau Y}$ is a directional quantile regression coefficient and $\boldsymbol \beta_{\tau X}$  is the usual quantile regression coefficient vector. 

For every direction $\boldsymbol u$ taking into consideration, its relative hyperplane as in Definition 2.1 characterize a region which connects the notions of Tukey depth and directional quantiles. We can say that each element $(\hat{a}_\tau, \hat{\boldsymbol b}_\tau, \hat{\boldsymbol \beta}_\tau)$ define an upper closed quantile halfspace
\begin{equation} \label{halfspace}
 H^+_{\tau \boldsymbol u} = H^+_{\tau \boldsymbol u} (\hat{a}_\tau, \hat{\boldsymbol b}_\tau, \hat{\boldsymbol \beta}_\tau) = \{ \boldsymbol y \in \mathbb{R}^k : \boldsymbol u^{'} \boldsymbol y \geq \hat{\boldsymbol b}_\tau \boldsymbol \Gamma^{'}_u \boldsymbol y + \boldsymbol x^{'} \hat{\boldsymbol \beta}_\tau  + \hat{a}_\tau \}
\end{equation}
and an analogous lower open quantile halfspace switching $\geq$ for $<$ in \eqref{halfspace}. Moreover, fixing $\tau$ we are able to define the $\tau$ quantile region $R(\tau)$ as 
\begin{equation} \label{tauRegion}
 R(\tau) = \bigcap_{{\boldsymbol u} \in \mathcal{S}^{k-1}} H_{\tau \boldsymbol u}^+,
\end{equation}
where the term $H_{\tau \boldsymbol u}^+$ might be stated slightly different, as $\cap\{ H_{\tau \boldsymbol u}^+ \}$, if the solution of the minimization problem is not unique. These quantile regions are important as they can be used to study the effect of the predictor variables in the conditional distribution of $\boldsymbol Y$. 

Moreover, these quantile regions are related to the Tukey depth definition described earlier in this section. \citet{hallin:10} proved that indeed $R(\tau) = D(\tau)$, under certain conditions. This is certainly noteworthy given that the computation of the Tukey depth has been a point of discussion in the literature \citep[see, for instance,][]{dyckerhoff2016}. Now due to this directional quantile approach one can obtain the Tukey depth region $D(\tau)$ through the analysis of $R(\tau)$. For an algorithm on how to compute those regions, see \citet{paindaveine2012}. The boundary of the quantile region is called quantile contour and usually it is the one information more appropriate to make comparisons, instead of the whole quantile region. One might be interested in comparing different quantile contours given different sets of the predictor variables, for example. Regarding the regression case for the computation of these quantile contours, \citet{hallin2017} argues that one needs severe assumptions in order for these quantities to be the depth contours of the conditional distribution of $\boldsymbol Y$ given $\boldsymbol X$. An alternative would be to consider the local bilinear quantile contour method proposed by \citet{hallin2015}. Despite that, here we will consider the boundary of the quantile regions defined in \eqref{tauRegion} obtained by the halfspaces defined in \eqref{halfspace}, as these can be considered an averaged version of the Tukey depth contour. Also, when we write conditional quantiles in this paper, we refer to the fact that these quantities are conditional on the covariate values. \citet{guggisberg2017} discusses this issue and the difference to the approach proposed by \citet{hallin2015}, indicating that the latter method could be named in fact conditional quantiles, instead of our approach. Regardless of this we continue with the former description, as we believe this does not cause confusion.

Finally, the minimization problem stated in \eqref{minization} can be solved using algorithms provided by the univariate quantile regression models. This is true considering that one may see this minimization as a regression problem of $Y_u = \boldsymbol u^{'} \boldsymbol Y$ on explanatory variables $\boldsymbol X$ and $\boldsymbol Y_u^\perp$ with an intercept term. The same path will be considered in the next section to define a Bayesian scheme to estimate these models, while also adding structured additive predictors and controlling for crossing quantiles.

\section{Bayesian estimation with structured additive predictors and noncrossing conditional quantiles}
\label{STADNCQR}

A Bayesian quantile regression model was first proposed by \citet{yu:01} and it considers in the likelihood the assumption of the asymmetric Laplace distribution. The association of quantile regression and this distribution was first noted by \citet{koenker:99}, with a likelihood ratio test for the quantile regression parameters. Let the density of a variable with asymmetric Laplace distribution, say $Y \sim AL(\mu, \sigma, \tau)$, as
\begin{equation} \label{density}
 f(y; \mu, \sigma, \tau) = \frac{\tau(1-\tau)}{\sigma} \exp \left\{ -\rho_\tau\left( \frac{y - \mu}{\sigma}\right) \right\},
\end{equation}
where $\mu \in \mathbb{R}$, $\sigma > 0$ and $\tau \in [0,1]$. Then the connection between these two ideas becomes clear as one can see that minimization in \eqref{minization} is equivalent to maximizing a likelihood given $n$ observations with density as in \eqref{density} for $Y_u$, replacing $\mu$ by a specific linear predictor including a directional axis $Y^\perp_u$. 

Moreover, a more efficient representation for this distribution was proposed by \citet{kozumi:11}, which allows for more extensions of these Bayesian quantile regression models. This mixture location-scale representation of this distribution can be stated as follows 
\begin{equation} \label{mixture}
 Y_i | v_i \sim N(\mu + \theta v_i, \psi^2 \sigma v_i), v_i \sim \mbox{Exp}(\sigma) 
 \Leftrightarrow Y \sim AL(\mu, \sigma, \tau),
\end{equation}
where $ \theta = (1-2\tau)/(\tau(1-\tau))$, $\psi^2 = 2/(\tau(1-\tau))$ and $\mbox{Exp}(\sigma)$ denotes the exponential distribution with mean $\sigma$. For the multiple-output response variable, one would have to make the assumption of the asymmetric Laplace distribution for the transformed variable $Y_u$ while adding the regression term $Y_u^\perp$. This was discussed by \citet{guggisberg2017}, where the author showed posterior consistency for this directional quantile approach. The technique was similar to the one considered in \citet{sriram:13}, where taking into consideration misspecified models, then one can still show how the posterior distribution still converges to the neighborhood of the true parameters. This is valid even when the true data generating process is not the asymmetric Laplace distribution, under certain conditions \citep[see][]{sriram:13}. This is also correct considering this multiple-output Bayesian quantile regression model according to \citet{guggisberg2017}. Similar to the single-output method, small sample sizes also do not present good coverage probabilities in our experience. In our applications though we consider only data sets with moderate sample sizes, for which we can have more confidence for consistency properties.

\begin{figure}[!ht]\centering
\includegraphics[width=6cm]{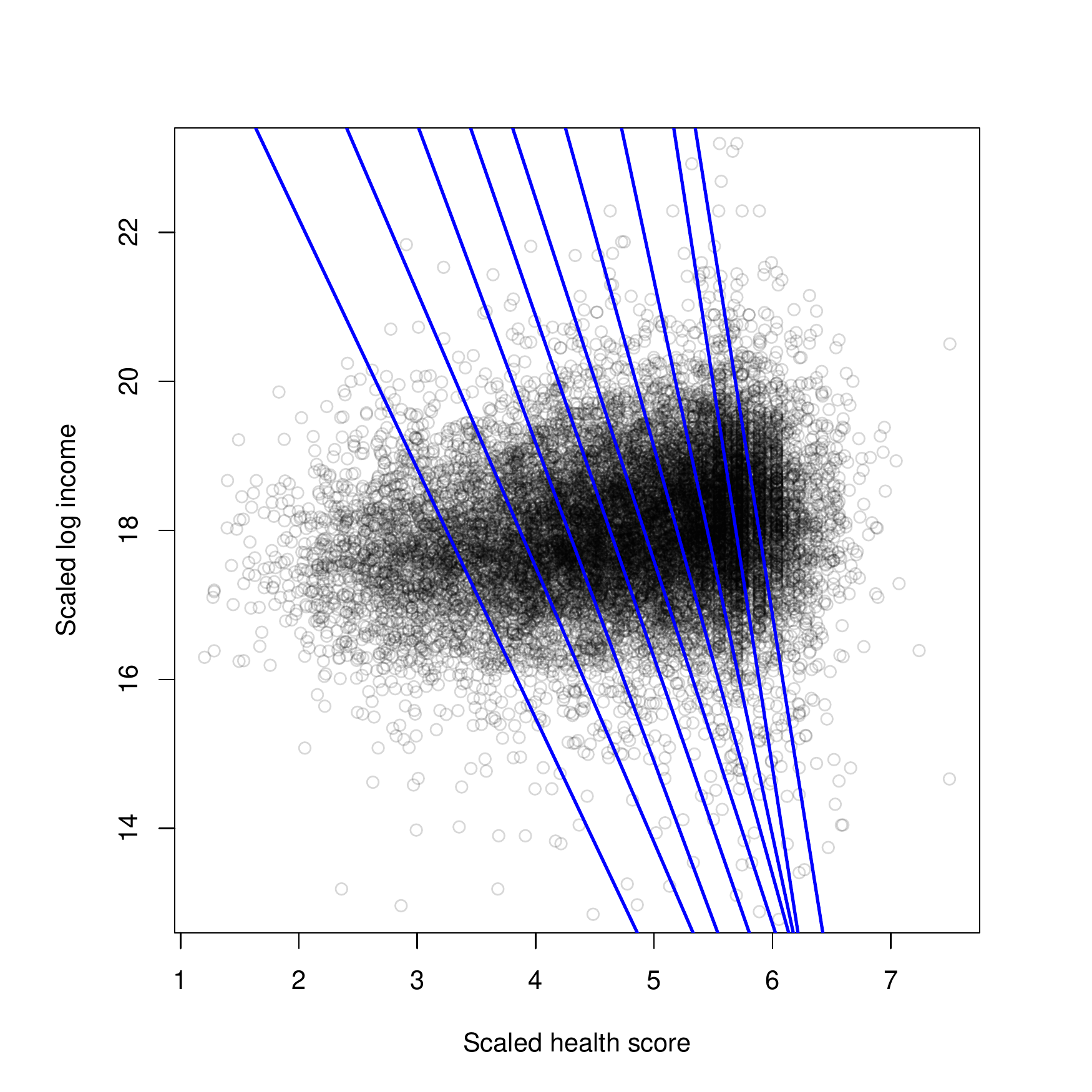}
\includegraphics[width=6cm]{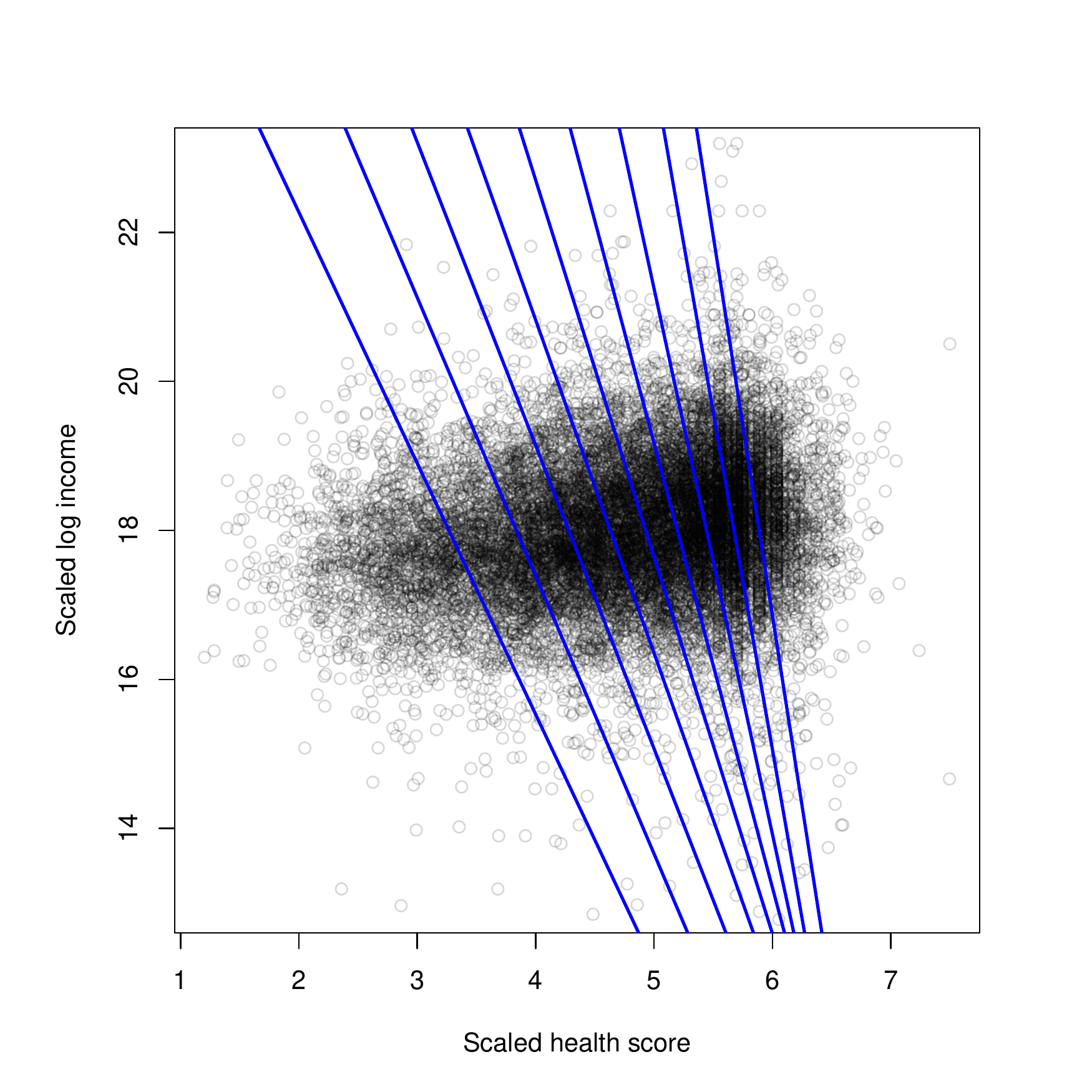}
\caption{\label{figureCrossings} Directional quantile hyperplanes for $\tau = 0.10, 0.20, \ldots, 0.90$ for $u = (3/\sqrt{10}, 1/\sqrt{10})$ before (left) and after the Gaussian process regression adjustment (right).}
\end{figure}

One problem which is present for quantile regression models for single-output response variables is crossing quantiles. This issue is important because it violates a simple probabilistic assumption, which states that the quantile function is nondecreasing for $\tau$. \citet{hallin:10} defines the multidimensional version of this problem when the quantile regions in \eqref{tauRegion} are nonnested. One possible solution for this difficulty is to use the methods proposed for single-output quantile regression models to fix crossing quantiles. In our case, this rearrangement could be made in the quantile hyperplanes $\lambda_\tau(\boldsymbol X)$. Though by definition these hyperplanes could cross each other, as pointed out by one referee, we argue in the next section that their rearrangement is sufficient to produce nested quantile regions. For instance, let us consider the application of Section~\ref{application}, where interest lies in analyzing inequality in both dimensions of income and health, considering data from Germany. If we take the direction $u = (3/\sqrt{10}, 1/\sqrt{10})$, then we would arrive in the plots portrayed in the left side of Figure~\ref{figureCrossings}, 
for the directional quantile hyperplanes at $\tau = 0.10, 0,20, \ldots, 0.90$, when we do not consider any predictor variable. One can notice that the obtained hyperplanes cross in a sparse region of the data. In the application section, we show that in the presence of covariates we arrive at nonnested quantile regions for some combination of variables. We propose then applying methods aimed at fixing crossing quantiles for single-output response variables for these directional quantile hyperplanes. Within the Bayesian framework, \citet{Rodrigues2017} proposed an adjustment based on a Gaussian process that considers a connection between all quantiles of interest through a correlation function. The same directional quantiles after this adjustment are depicted on the right side of Figure~\ref{figureCrossings}. We show empirically that this modification is able to produce nested quantile regions indeed.

\subsection{Gaussian process regression adjustment for all directions}

Let $\boldsymbol \tau = \{\tau_1, \ldots, \tau_L \}$ be the set of quantiles of interest for each direction. Then, if we assume the asymmetric Laplace distribution for the transformed variable $Y_u$ we can use the information from neighboring quantiles to correct the crossing problem using the approach proposed by \citet{Rodrigues2017}. For that, first remember that we can write the conditional quantile function of $Y_u | X \sim AL(\mu, \sigma, \tau)$ as a function of the parameter vector $\theta = (\mu, \sigma, \tau)$ \citep[see][]{yu:05} as 
\begin{align*}
 Q_{Y_u}(p | \theta) &= F^{-1}(p; \theta) \\
 &= \begin{cases}
     \mu + \displaystyle \frac{\sigma}{1-\tau} \log \left( \frac{p}{\tau} \right), & \mbox{if } 0 \leq p \leq \tau \\
     \mu - \displaystyle  \frac{\sigma}{\tau} \log \left( \frac{1-p}{1-\tau} \right), & \mbox{if } \tau \leq p \leq 1 \\
    \end{cases},
\end{align*}
which can be seen as the induced quantile function for all $p \in (0, 1)$ given a particular fixed $\tau$. If we substitute $\mu$ by its linear predictor, for instance, and take every posterior sample $t$ obtained in MCMC procedure, then we can have a sequence of induced conditional quantile values for each fixed $\tau$ and each direction $\boldsymbol u$, $Q_{Y_u}^{(t)}(p | \theta)$. An important result from \citet{Rodrigues2017} states that one can consider only the mean of the $p$th quantile from each model $\tau \in \boldsymbol \tau$, as 
\[
 \hat{Q}_s(p | \theta) = \frac{1}{T} \sum_{t=1}^T Q^{(t)}(p| \theta),
\]
where $T$ is the number of posterior samples. Then to smooth the information between quantiles, one can assume a Gaussian process to these induced posterior means, as 
\begin{align} \label{gaussianprocess}
 \hat{Q}_s(p | \theta) &= g(\tau) + \epsilon, \quad \mbox{with } g(\tau) \sim GP(0, K), \\
 \epsilon &\sim N(0, \Sigma), \nonumber
\end{align}
where $\Sigma$ is a $L \times L$ covariance diagonal matrix, with entries $\sigma^2(p | \tau)$, posterior variances of the corresponding $Q^{(t)}(p| \theta)$, and $K$ is a $L \times L$ covariance matrix with entries calculated by the squared exponential kernel 
\[
 k(\tau, \tau^{'}) = \sigma^2_k \exp \left\{ - \frac{1}{2 b^2} (\tau - \tau^{'})^2 \right\},
\]
which controls the correlation between the estimated quantiles, where $b$ is the bandwidth parameter and $\sigma^2_k$ is a variance hyperparameter of the prior. The authors of the method suggests that $\sigma^2_k$ can be set to 100, for example, without causing performance issues and also resulting in a noninformative prior. Though one must be careful with larger values for this parameter, as it might create computational problems while inverting matrices involved in the process. 

The Gaussian process in \eqref{gaussianprocess} allows one to model the correlation structure between the quantiles of interest for a given direction, making it possible to use the information from all quantiles to obtain a final quantile estimate. The final $p$th   estimate is then the adjusted posterior mean, $\hat{Q}_a(p | \theta)$, which is obtained from the posterior predictive distribution at $\tau = p$. Let $g(\tau = p) \sim N(\mu_*, \sigma^2_*)$ be the posterior predictive distribution, then we have the following
\begin{align*}
 \mu_* &= \hat{Q}_a(p | \theta) = \sum_{l=1}^L \omega_l \hat{Q}_s(p | \theta_l) \mbox{ and } \\
 \sigma^2_* &= \sigma^{{2}^{'}} + \sigma^2(p | \tau = p),
\end{align*}
where $\sigma^{{2}^{'}} = k(p, p) - {\boldsymbol W} {\boldsymbol K(., p)}$ and $\boldsymbol W = \boldsymbol K(., p)^T(\boldsymbol K + \boldsymbol \Sigma)^{-1}$, and $\omega_l$ is a term of the row vector of weights $\boldsymbol W$.

\citet{Rodrigues2017} show that for any $\tau_1 \leq \cdots \leq \tau_L$, it always exist a bandwidth $b$ such that $\hat{Q}_a(\theta_1 | \theta) \leq \cdots \leq \hat{Q}_a(\theta_L | \theta)$, solving the problem of noncrossing quantiles, which might affect quantiles when these are estimated separately. The authors also proved that posterior consistency is achieved, if the initial model is consistent. Given these results, they suggest an algorithm, for which we suggest a modification, given the ordered vector of directions $\boldsymbol U = \{u_1, \ldots, u_D\}$.  
\begin{description}
 \item[1.] Fit $L$ separate quantile regression models, $\boldsymbol \tau = \{\tau_1, \ldots, \tau_L \}$;
 \item[2.] Obtain induced quantile posterior means $\hat{Q}_s(\boldsymbol \tau | \theta)$, for all $X$ and $\boldsymbol \tau = \{\tau_1, \ldots, \tau_L \}$;
 \item[3a.] For the first direction $u_1$, initialize the algorithm with $b \approx 0$ and keep increasing $b$ by appropriate values until there are no quantile estimates crossing. Do this check for $X$ positioned in the hull of the sample space and $\boldsymbol \tau = \{\tau_1, \ldots, \tau_L \}$.
 \item[3b.] For $u_k, \, 1 < k \leq D$, initialize the algorithm with $b = b_{k-1}$, the bandwidth of the previous direction and decide whether to increase or decrease $b$, given the presence or absence of crossing quantiles, respectively.
\end{description}

For our directional model, the only difference between our approach and the one proposed by \citet{Rodrigues2017} is the possibility of using the bandwidth of neighboring directions, which borrows the information along the different directions and could improve the performance of the algorithm. This adjustment for each direction $u$ allows us to define the quantile region as in \eqref{tauRegion} based on halfspaces determined after the Gaussian process regression adjustment, as
\begin{equation} \label{tauRegionAd}
 R_a(\tau) = \bigcap_{{\boldsymbol u} \in \mathcal{S}^{k-1}} H_{\tau \boldsymbol u_a}^+,
\end{equation}
where the regression adjustment is performed for every ${\boldsymbol u} \in \mathcal{S}^{k-1}$. Then by sequentially applying the results obtained by \citet{Rodrigues2017}, one can check that the $\tau$-quantile regions are nested 
$$R_a(\tau_1) \subseteq R_a(\tau_2),$$ 
for $\tau_1 < \tau_2$. In practice, one defines a set of directions $u_1, \ldots, u_D$, then for every direction the Gaussian process regression adjustment is used in case of quantiles crossing. Therefore, quantile regions based on these new estimates will preserve the nesting property.

As correctly pointed out by one referee, it should be noted that this Gaussian process regression adjustment for each direction is not a necessary condition to obtain nested quantile regions, though it is a sufficient one. Taking into account that often this adjustment is needed in sparse regions of the covariate space, which is usually also the case for single-output models, one can expect that this modification will not lead to false statistical properties. One of the properties regarding these directional quantile regression is defined by \citet{hallin:10} as a subgradient condition. This states that
\begin{equation} \label{subgradient}
 P(\boldsymbol Y \in H_{\tau \boldsymbol u}^-) - \tau = 0, 
\end{equation}
which relates this directional multiple-output method to the same probability nature of single-output quantiles. In order for this Gaussian process adjustment to be valid, one needs to check whether this is still valid after the adjustment, i.e, whether $P(Y \in H_{\tau \boldsymbol u_a}^-) = \tau$ is still true. We suggest that this check should be made whenever one considers this approach and in fact this can be easily verified in the data. We show these results in our application.

\subsection{Structured additive predictors} 

While the approach proposed by \citet{hallin:10} provides a way of analyzing these directional quantiles of response variables of dimension greater than one, extensions to allow for more different type of predictors, e.g., nonlinear functions, spatial or even random effects, might not be so direct. This is due to the linear programming algorithm, which is necessary for estimation, and related inferential procedures attached to this technique. Yet, it would be interesting to not only model this type of multiple-output response, but also to add some flexibility in the structure of the predictor variables. \citet{hallin2015} provide the option of estimating nonlinear quantile contours, but their approach would not be able to consider also spatial effects. For instance, when modeling data about income and health inequalities in the next section, we could be interested in adding not only nonlinear effects, but also spatial terms.

Taking that into consideration, we follow the proposal of \citet{waldmann2013} for Bayesian semiparametric quantile regression models. For presentation purposes, here we make the necessary adjustments for our multiple-output response variable scenario. In this case, we assume that given a sample of $n$ observations of $\boldsymbol Y$ each $Y_{iu} \sim AL(\eta_{iu}, \sigma_u, \tau)$. Then we can write the predictor for the location parameter as 
\begin{equation} \label{genPredictor}
\eta_{iu} = \sum_{j=1}^q f_{ju}(z_i) + \boldsymbol{x}_i^{'} \boldsymbol \beta_\tau + \boldsymbol y_{iu}^\perp \boldsymbol b_\tau, 
\end{equation}
where function $f_j$ in \eqref{genPredictor} can be defined according to specific cases. For example, a nonlinear function for a continuous variable could be considered using $f_j(z_i) = f(z_i)$, where $f$ could be approximated with p-splines or other basis functions; spatial effects could be studied through a proper definition of $f_j$, as well. Taking into account the basis definition in \eqref{basisDefinition}, then one can write the predictor vector using a matrix notation, 
\begin{equation*}
 \boldsymbol \eta_u = \boldsymbol Z_1 \boldsymbol \gamma_{1\tau} + \cdots + \boldsymbol Z_q \boldsymbol \gamma_{q\tau} + X \boldsymbol \beta_\tau + Y_u^\perp \boldsymbol b_\tau,
\end{equation*}
where we consider suitable basis expansions for each $\boldsymbol Z_j$ and $\boldsymbol \gamma_{j\tau}$ contain its relative coefficients, for $j = 1, \ldots, q$.

If we consider the mixture defined in \eqref{mixture}, we have the following 
\begin{equation} \label{transformedModel}
 Y_u | \boldsymbol b_\tau, \boldsymbol \beta_\tau, \boldsymbol \gamma_\tau, \sigma, \boldsymbol v \sim N(\boldsymbol \eta_u + \theta \boldsymbol v, \psi^2 \sigma \boldsymbol V),
\end{equation}
where $\boldsymbol \gamma_\tau = (\boldsymbol \gamma_{1\tau}, \ldots, \boldsymbol \gamma_{q\tau})$, $\boldsymbol V = \mbox{diag}(v_1, \ldots, v_n) $ and $\boldsymbol \beta_\tau$ contains an intercept. Then one needs to define priors distributions for $\xi = (\boldsymbol b_\tau, \boldsymbol \beta_\tau, \boldsymbol \gamma_\tau, \sigma)$ to complete the specification of the model. \citet{guggisberg2017} discusses prior elicitation for $\boldsymbol b_\tau$ and an intercept term $a_\tau$, in which the author relates this prior to Tukey depth of the data. \citet{waldmann2013} show how one can setup the priors for $\boldsymbol \gamma_\tau$ accordingly for different type of predictors. Different from \citet{guggisberg2017} we do not fix the $\sigma$ parameter to 1, instead we consider a gamma distribution as in \citet{waldmann2013}. In fact, all prior distributions and related full posterior conditional distributions are similar to the ones presented in the latter, with the only difference being the transformed variable $Y_u$ of \eqref{transformedModel}. Due to this, we refer to \citet{waldmann2013} to these results.

It is important to mention that the addition of the additive terms to the estimation does not change the innovation regarding noncrossing quantiles. What might increase is only the computational cost, as these additive should increase the number of coefficients, which increases the dimension of the Gaussian process considered in \eqref{gaussianprocess}. In this case, the search for a bandwidth $b$ that fixes the quantile crossing problem might become more troublesome. 

\section{Applications}
\label{application} 

\subsection{Inequality data in Germany}

In order to illustrate how this directional approach can shed new light to the analysis of quantile regression models, we consider data from the Socio Economic Panel \citep{SOEP} collected in Germany in the year of 2012, which was also analyzed by \citet{silbersdorff2018}. For this exercise, we are interested in two dimensions of inequality in the population, namely health and income. For the former, we use a standardized health score which gives a good approximation of the physical well being of each person, while for the latter we consider the logarithm of the income of the household. In their analysis, the authors considered income as one of the explanatory variables to try to correlate these two variables, while controlling for other variables. Here we are interested in checking how other variables such as age, education, marital status might affect jointly these two response variables, considering this directional Bayesian quantile regression concept.

In order to explain the conditional distribution of these two variables, we select a few variables to estimate these directional bivariate quantile regression models. These are age, sex, educational level (edu: elementary education or less [1], secondary education [2], higher vocational training [3], completed higher education [4]), family status (FamStat: married [1], separated or divorced [2], single [3] or widowed [4]), and a dummy variable to differentiate the West from the East part of Germany (EW). Both responses variables were scaled, i.e., divided to its standard deviation, in order to make comparisons easier. The following model was then estimated for 99 equally spaced vectors in the unit circle in $\mathbb{R}^2$ starting with $u = (1,0)$, where the $x$ coordinate is the health score and the $y$ coordinate is log income,
{\small 
\begin{align*}
 Q_{\boldsymbol Y_{u}}(&\tau|X) = \beta_{0\tau}^u + \beta_{1\tau}^u Y_u^\perp + f_\tau^u(\mbox{age}) + \beta_{1\tau}^u \mbox{edu2}  + \beta_{2 \tau}^u \mbox{edu3} + \beta_{3\tau}^u \mbox{edu4}\\
 & + \beta_{4\tau}^u \mbox{FamStat2} + \beta_{5\tau}^u \mbox{FamStat3} + \beta_{6\tau}^u \mbox{FamStat4} + \beta_{7\tau}^u \mbox{EW} + \beta_{8\tau}^u \mbox{sex},
\end{align*}} 
where the indexes $u$ and $\tau$ are used to indicate the dependence of the regression parameters on the direction and the quantile for every model. For variable age is modeled its effect through a nonlinear function $f_\tau^u(.)$, which for this example we considered cubic P-splines with 20 equidistant knots. Estimation was done with the BayesX software \citep{belitz2015} calling its routines through a wrapper function in the software R. 

The prior distributions used in this application are mainly noninformative distributions. For the nonlinear effect, the coefficients {$\boldsymbol \gamma$} were given a second order random walk, i.e., 
\[
 \gamma_k \sim N\left( 2\gamma_{k-1} - \gamma_{k-1}, \frac{1}{\nu} \right) \quad k = 3, \ldots, K,
\]
where $\nu \sim Ga(a, b)$, with $a = b = 0.001$. For the fixed effects, a normal prior distribution is defined centred at zero, with precision parameter $\zeta$, where $\zeta $ is given a gamma distribution, with hyperparameters equal to 0.001 as well. The latent variables $v_i$ have an exponential distribution with parameter $\sigma$, for which is given a gamma prior distribution with $a = b = 0.001$. We have tried different values for the hyperparameters in this application, but the posterior estimates did not present any meaningful variation. All results were obtained with chain size of 55000 samples, after discarding the first 5000 draws and recording every 50th value. All estimates are based on the posterior means. We considered different values for $\tau$ to showcase the different features of the model and each figure in this application have the values of $\tau$ that were used.

\begin{figure}[!ht]\centering

\includegraphics[width=7cm]{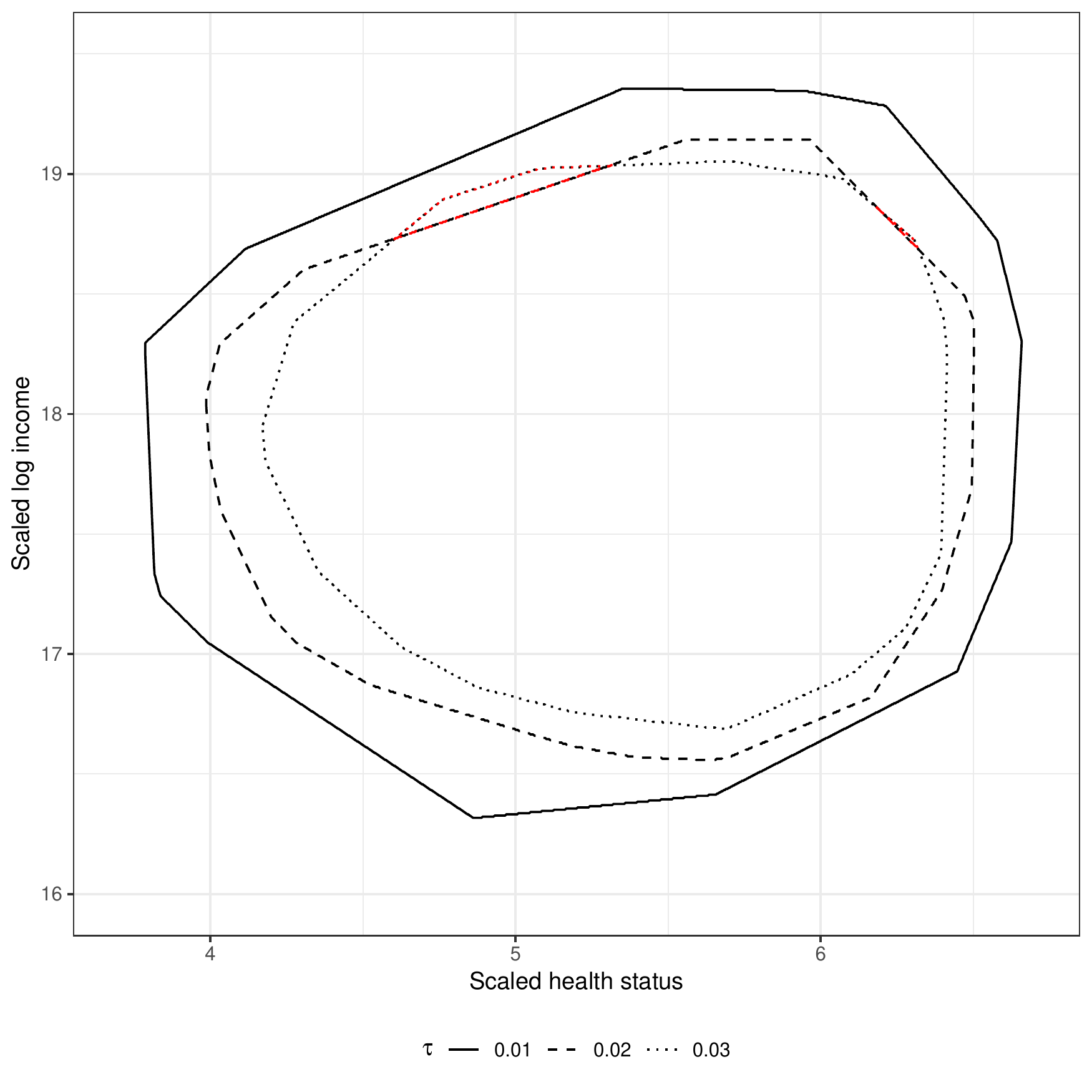}
\includegraphics[width=7cm]{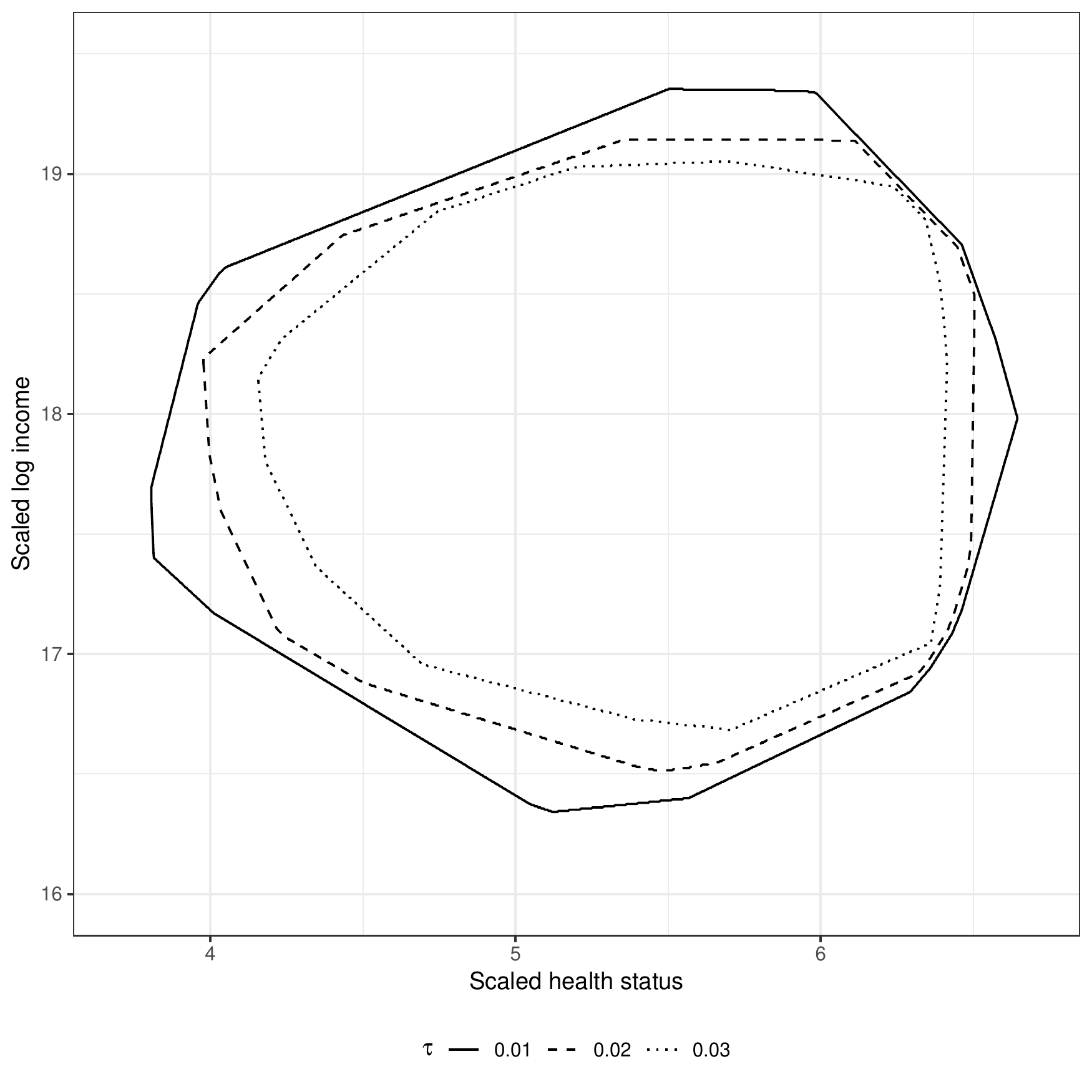}
\caption{\label{quantile_crossing} Quantile contour before and after using the Gaussian process regression adjustment for $\tau = 0.01, 0.02 \mbox{ and } 0.03$. The red lines indicate where the crossing happens in the plot on the left side.}
\end{figure}

When we analyze this directional method, a first challenge is to visualize all the results, taking into considerations all these chosen directions. Possibly the easier way to compare estimates is calculating conditional quantile contours for different combinations of the covariates and different quantiles. As we discussed previously about the crossing problem for quantile regression models, here we select a combination in this application which displays the crossing problem. For instance, in Figure~\ref{quantile_crossing}, one can see the estimated quantile contours for a 20 years old widowed man with vocational training from the Western part of Germany, for $\tau = 0.01, 0.02 \mbox{ and } 0.03$. The plot on the left side shows the crossing of the quantile contours in the direction of lower health status and higher income
and also for higher income and higher health status, for $\tau = 0.02$ and $\tau = 0.03$. The locations where crossing of conditional quantiles happens are highlighted in red. And the plot on the right side presents the estimates for the same quantiles, after using the Gaussian process regression adjustment for each direction estimate. In this second plot, the problem of crossing vanishes as proposed by this method. One should also take notice that this issue takes place in distinct directions. 

Regarding the subgradient condition in \eqref{subgradient}, we check these values for this application before and after the Gaussian process regression adjustment. We obtained the mean over the different and the results are presented in Table~\ref{tableSubgradient}. First, given that these are very extreme quantiles, it is not surprising that the subgradient conditions are not exactly met. Though one can certainly notice that for quantiles 0.02 and 0.03, their values actually improve after the adjustment. Overall one cannot notice a big difference between before and after the Gaussian process regression adjustment, which should just emphasize that this modification does not lead to a statistical dilemma.

\begin{table}[!htb]
\centering
\caption{Estimates for the subgradient condition, averaging over the different directions, in the application before and after the Gaussian process regression adjustment.}
\label{tableSubgradient}
\begin{tabular}{ccc}
\hline
\hline
 $\tau$ & Before & After \\
  \hline
0.01 & 0.0170 & 0.0208 \\
0.02 & 0.0316 & 0.0303 \\
0.03 & 0.0447 & 0.0404 \\
\hline
\hline
\end{tabular}
\end{table}

For this application, the bandwidths obtained to fix the crossing probem were on average equal to 139.41, but this value is influenced by the fact that in several directions there are no crossing quantiles. \citet{Rodrigues2017} recommend some caution for extreme quantiles and large values for the bandwidth, which could lead to biased estimates. Here, the largest value in one of the directions was equal 1200.50, which could be a cause for concern. Given the previously mentioned subgradient conditions for even these extreme quantiles, we consider that these adjustments still give a proper representation of the conditional quantiles.

\begin{figure}[!ht]\centering
\includegraphics[width=0.75\textwidth]{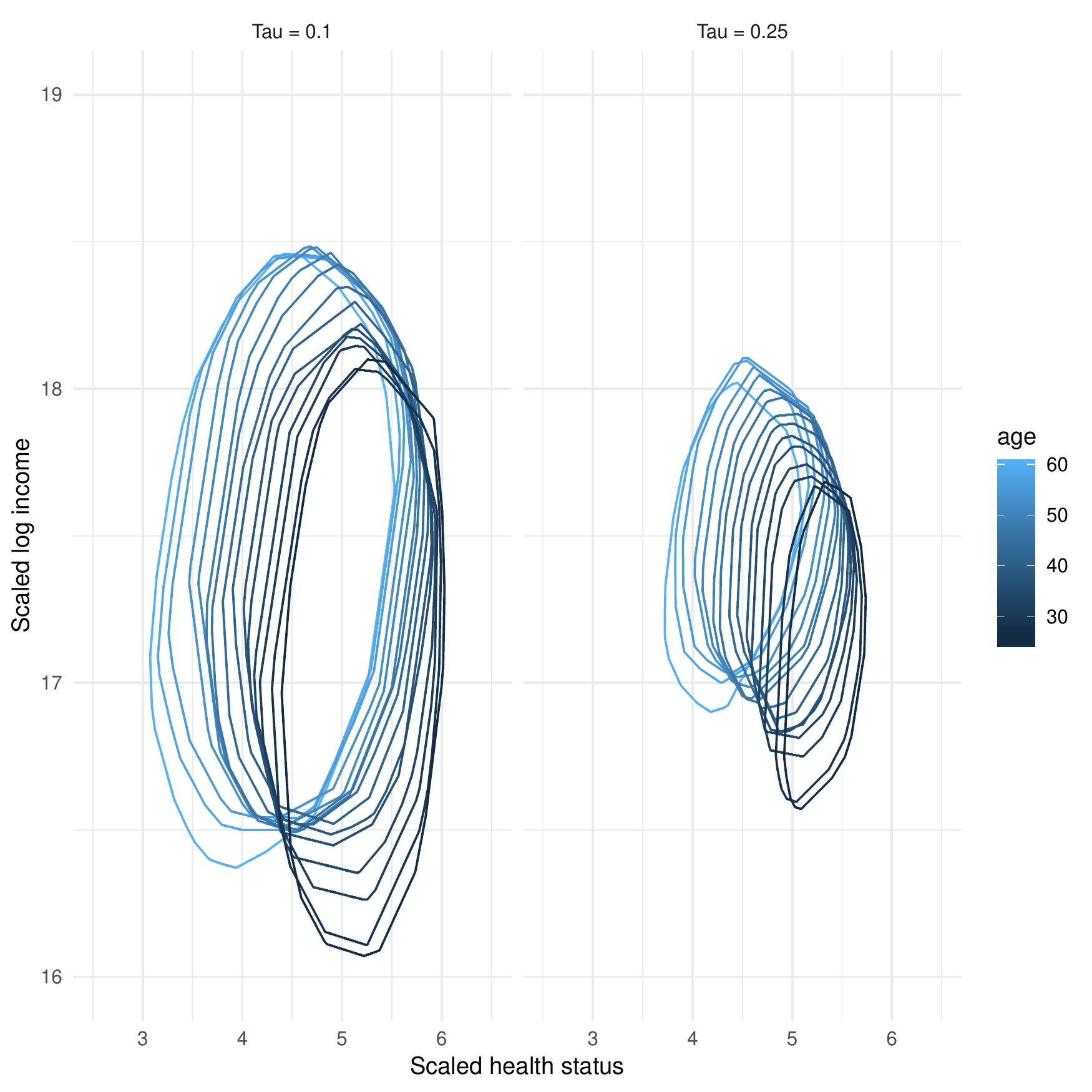}
\caption{\label{nonlinear_age} Quantile contours to show the nonlinear effect of age considering a sequence of values from 25 to 60 years old for every 2.5 years, and $\tau \in \{0.1, 0.25\}$}
\end{figure}

The nonlinear effect of age assumed in this application can also be displayed with the help of quantile contours. In this case, we plot the values for the quantile contours for values of age between 25 and 60 years old, in intervals of 2.5 years. The result is shown in Figure~\ref{nonlinear_age} for $\tau = 0.1, 0.25$. Though the clear nonlinear behavior is straightforward to be identified checking the estimated curve as function of age for each direction, one can still observe the variation of the quantile contours is definitely not linear. Overall, one can say that the effect of age is increasing for the income dimension, while having a negative impact on the health income. Additionally, variance for older ages seems to be greater for quantile 0.1.

\begin{figure}[!ht]\centering

\includegraphics[width=5.25cm]{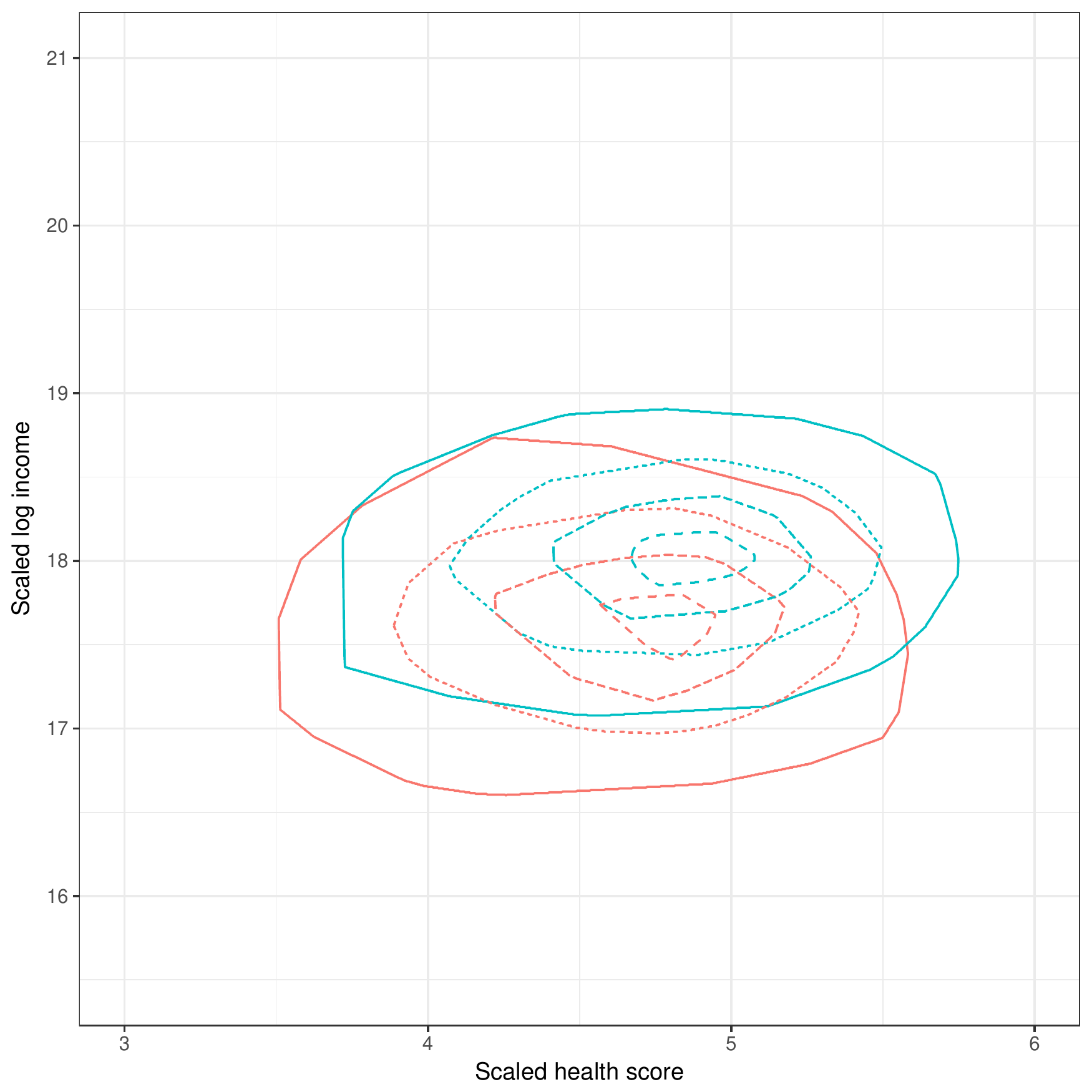}
\includegraphics[width=5.25cm]{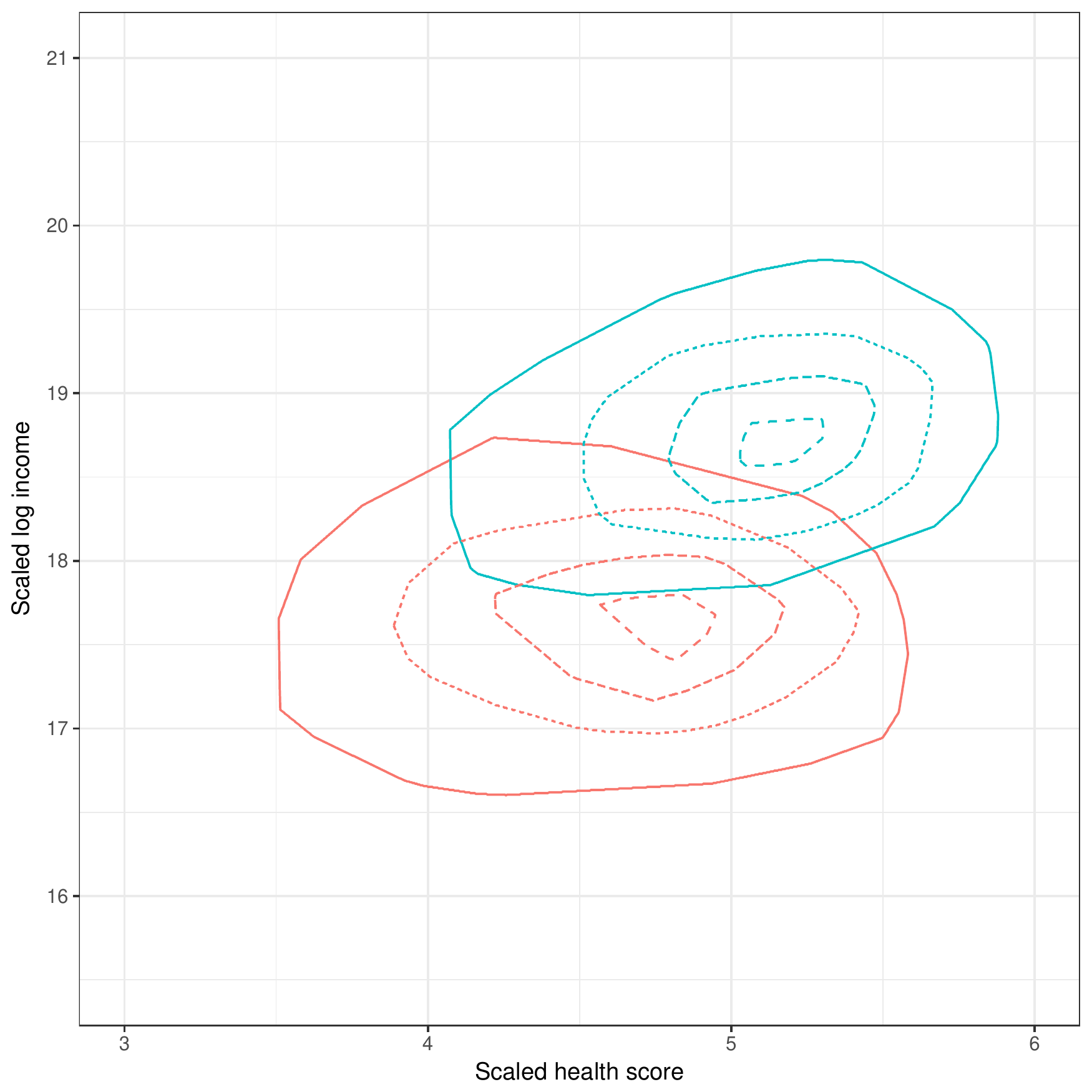}
\includegraphics[width=5.25cm]{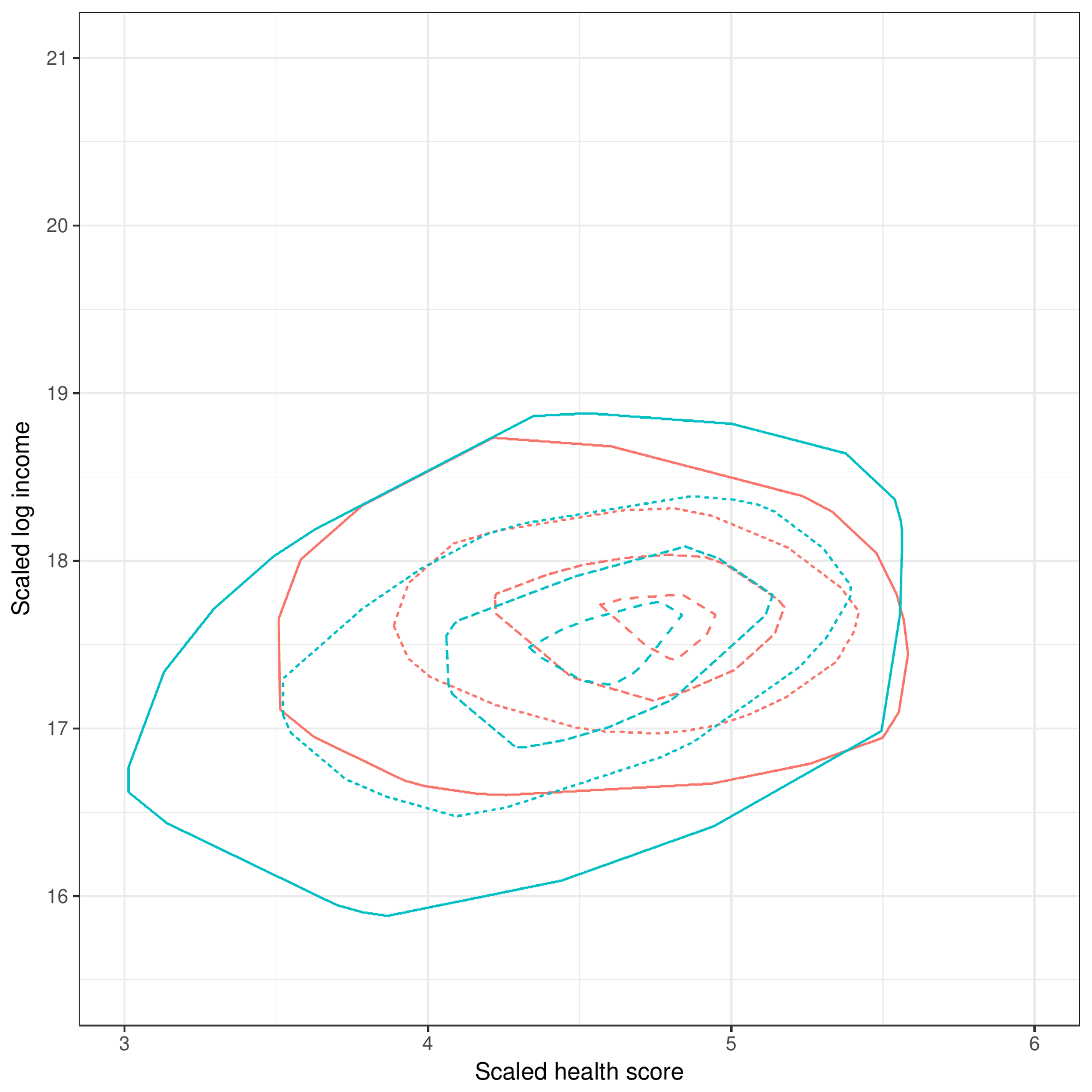}
\includegraphics[width=5.25cm]{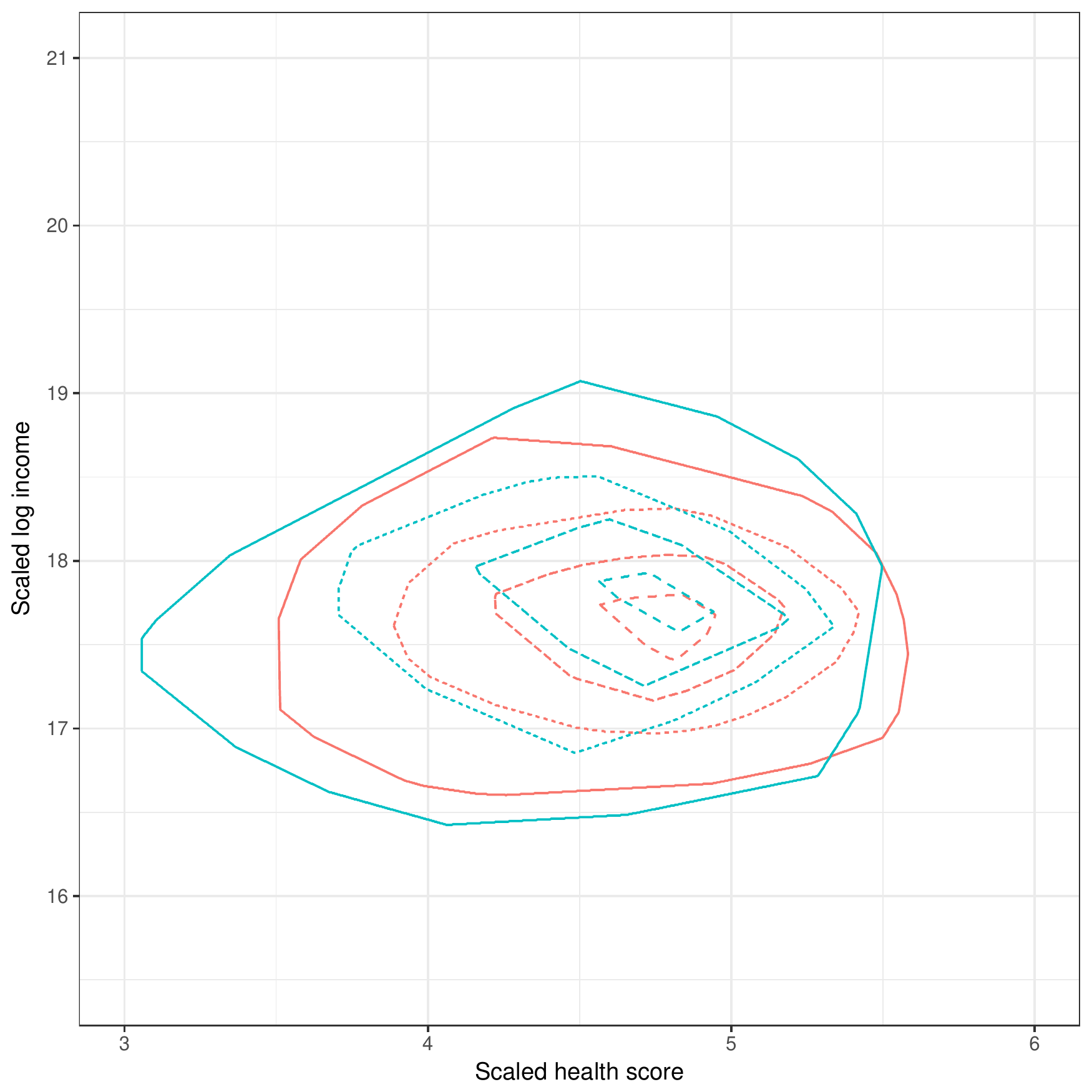}
\includegraphics[width=5.25cm]{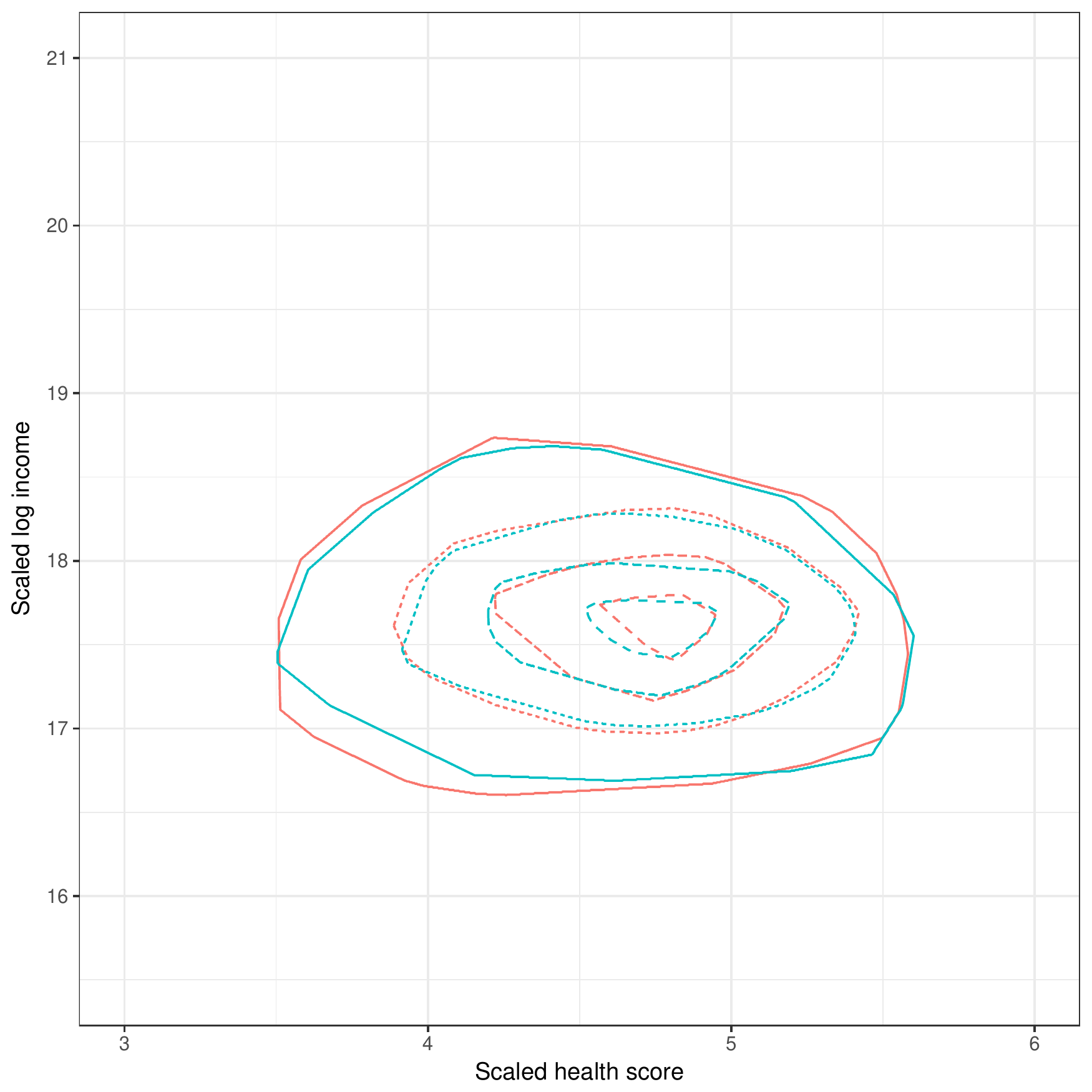}
\includegraphics[width=5.25cm]{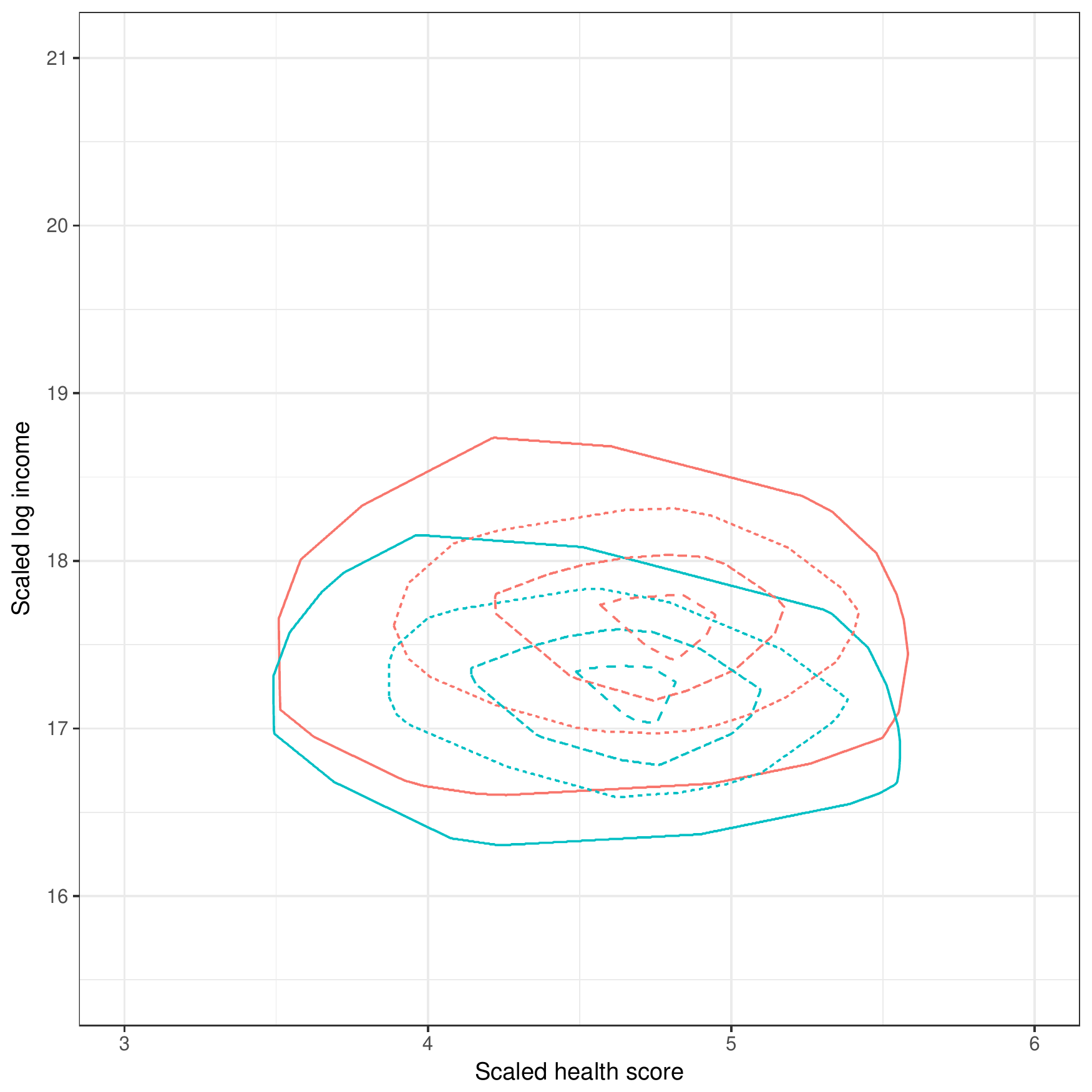}
\caption{\label{comp_2D_effect} Quantile contours for $\tau = 0.1, 0.2, 0.3, 0.4$, considering average values for variable age. First row, comparison with quantile contours between different levels of education: elementary education or less (blue),  secondary education (red left), completed higher education (red right). Second row, comparison with quantile contours between different levels of family status:  married (blue), separated or divorced (red left), widowed (red right). Third row, left side: comparison between men (red) and woman (blue). Third row, right side: comparison between West part of Germany (red) and East part of Germany (blue). The $\tau = 0.1$ quantile contour is the innermost in the figure, while $\tau = 0.4$ is the outermost for each scenario.}
\end{figure}

For instance, in the first row of Figure~\ref{comp_2D_effect}, we can compare the quantile contours of distinct levels of education. We have that the contrast between the elementary school or less is much stronger compared to completed higher education than compared to secondary education. While one can definitely notice the differences in the income dimension between the Tukey depths in the first plot, the difference in the health dimension is not so stark. A different result we get when we analyze the second plot concerning education levels. When we consider people with higher education then the differences to people with elementary education or less are evident in both dimensions for all quantiles. 

Moreover, we can also study the Tukey depth divergences along the marital status variable, as shown in the second row of Figure~\ref{comp_2D_effect}. In this case, we are interested in checking similarities between married, separated or divorced and widowed. The first category is the reference, with which we make the comparisons regarding the other two levels of the variable. The dissimilarities married and separated or divorced are more clear in the directions between $u_1 = (-1, 0)$ and $u_2 = (0, -1)$, which means for people with lower income and lower health status. As for widowed people, it is a more complicated to specify a kind of pattern of differences in regards to the reference levels, as that different directions or quantiles show different ordering for this pair of values for the predictor variable. 
 
Additionally, on one hand we can say that the differences between men and women are not as evident, given the last row of Figure~\ref{comp_2D_effect}. On the other hand, there are clear distinctions between East and West of Germany. Though those differences are also more evident in the income dimension. Overall, for different combination of predictor variables, we are able to identify different type of divergences between the quantile contours. This shows how this approach is really able to discuss in more detail the conditional inequalities of income and health status jointly given a set of predictor variables. This certainly helps in a considerable way when one wants to have a more complete picture of conditional distribution of these two response variables.

\subsection{Application to exam scores in Brazil}

For a second application, we use data from the Brazilian High School National Exam (ENEM, in Portuguese), from 2017, which is available at \url{http://portal.inep.gov.br/web/guest/microdados}. This exam is specially important because it is used for selecting students for admission in public universities in Brazil and it gives the scores for each student for different disciplines, such as natural sciences, languages or mathematics. This allows the analysis of the multivariate conditional distribution of these scores given other interesting covariates, such as sex, income and education levels of the parents, high school enrollment (private/public), among others. For this illustration, we have selected only students from the state of S\~ao Paulo, which were present for all days of the exam, from public and private schools, who had a positive score for all disciplines and were between age of 16 and 21 years old. After applying these filters we have 19912 observations.

\begin{figure}[!ht]\centering

\includegraphics[width=0.3\textwidth]{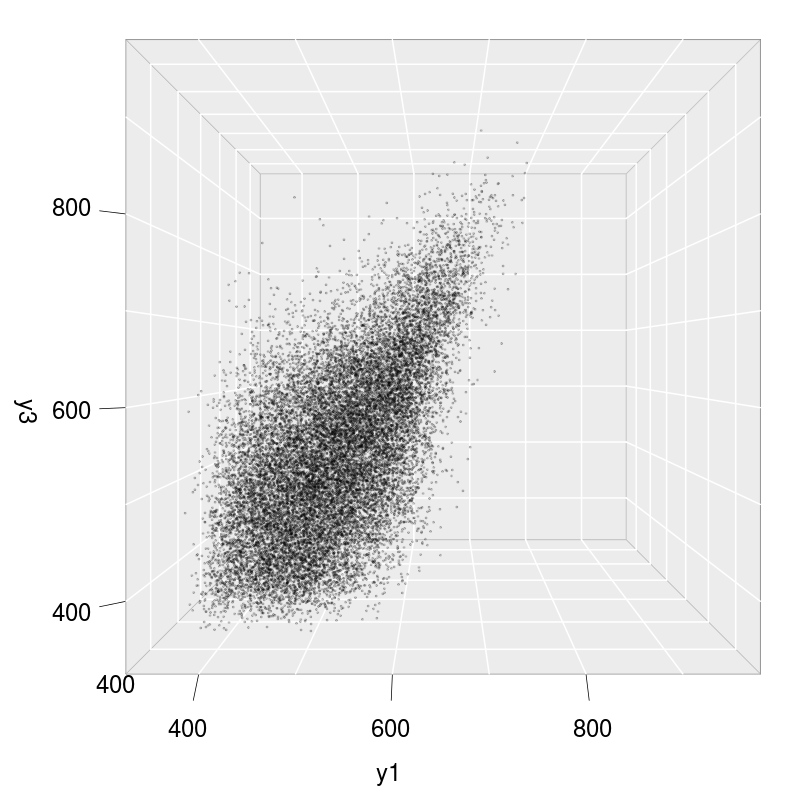}
\includegraphics[width=0.3\textwidth]{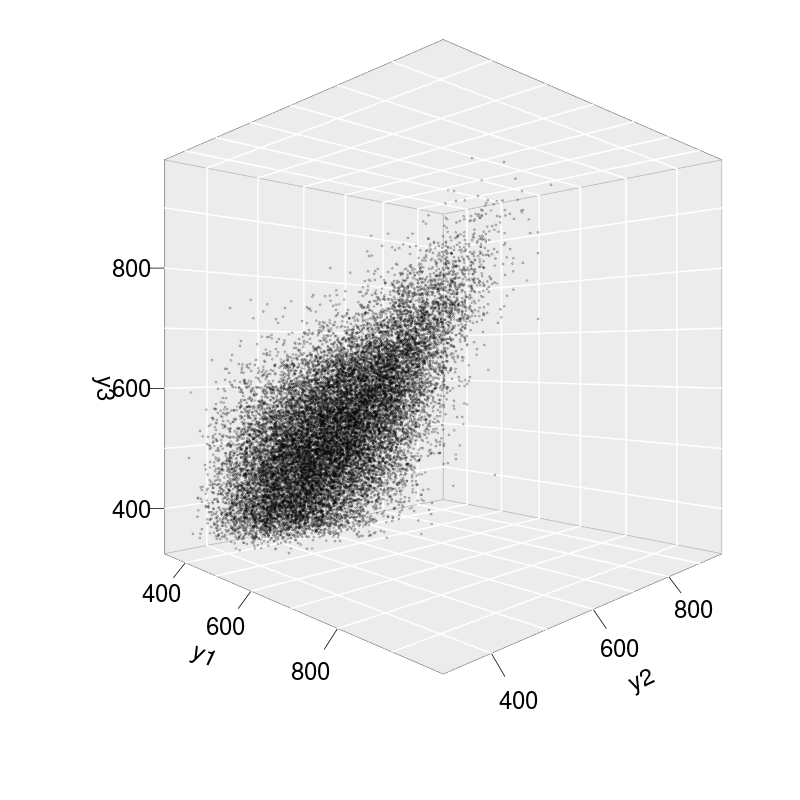}
\includegraphics[width=0.3\textwidth]{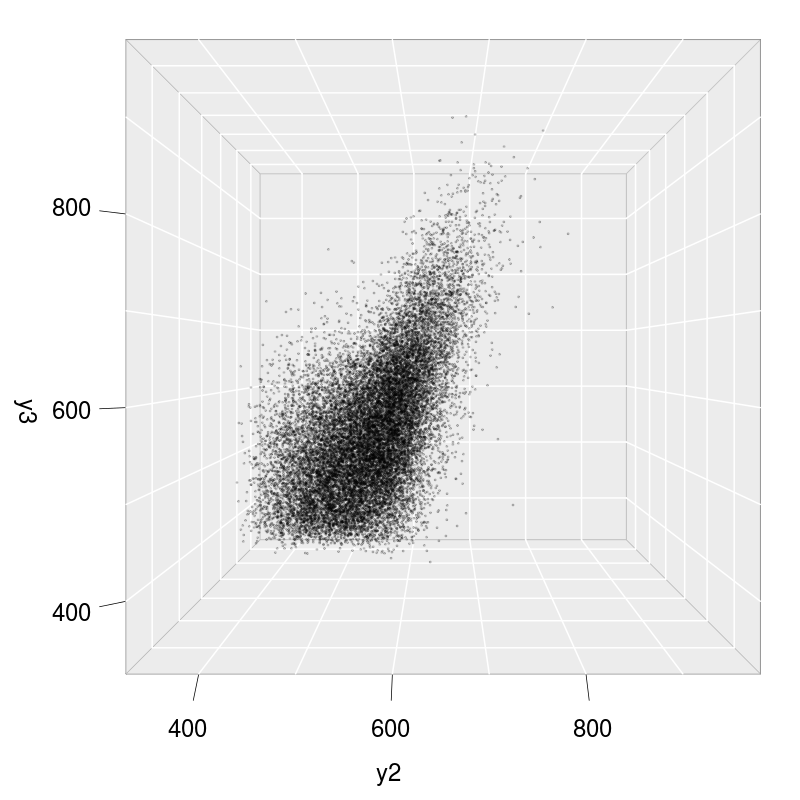}
\caption{\label{3D_data} Scatterplots of the multiple-output response variable, which is the score in the Brazilian High School National Exam for three different disciplines: $y_1$, natural sciences ; $y_2$, human sciences ; $y_3$, mathematics.}
\end{figure}

For our response variable, we use the scores in three different subjects: $y_1$, natural sciences ; $y_2$, human sciences ; $y_3$, mathematics. Our multiple output response is then $Y = (Y_1, Y_2, Y_3)$ and one can see a 3D scatterplot of the data in Figure~\ref{3D_data}. Given its 3 dimensional nature of the data, it becomes harder to visualize the whole variation in space. Due to this, for all plots in this subsection, we show the same plot in three ways, rotating the plot to improve the analysis. We consider the angles $\theta$ as the azimuthal direction and $\phi$ as the colatitude, fixing $\phi = 10^{\circ}$ and using values for $\theta = \{ 0^{\circ}, 45^{\circ}, 90^{\circ} \}$.

For the covariates in this application we consider sex ($x_1$: 0 = male, 1 = female), race ($x_2$: 0 = non-white, 1 = white), type of school ($x_3$: 0 = public, 1 = private), father\footnote{Here it was possible to answer the questionnaire based on the education of the man in charge of the person taking the exam. The same applies for mother education.} education ($x_4$: 0 = with at least university degree, 1 = otherwise), mother education ($x_5$: 0 = with at least university degree, 1 = otherwise), income level ($x_6$: 0 = household income less than R\$2811, 1 = otherwise), score on writing test ($x_7$), and score on languages test ($x_8$). It was considered 512 directions in the unit ball $S^2$ and a plot with all the directions considered can be seen in the supplementary material. The following model was then estimated for $\tau = 0.05, 0.10, \mbox{ and } 0.20$,
 {\small 
\begin{align*}
 Q_{\boldsymbol Y_{u}}(&\tau|X) = \beta_{0\tau}^u + \gamma_{1\tau}^u Y_u^\perp + \beta_{1\tau}^u x_1  + \beta_{2 \tau}^u x_2 + \beta_{3\tau}^u x_3 + \beta_{4\tau}^u x_4 + \beta_{5\tau}^u x_5 + \beta_{6\tau}^u x_6 + \beta_{7\tau}^u x_7 + ,
\end{align*}} 
for every $u$ direction. All parameters estimates were obtained from the posterior means for each parameter after running the MCMC chains for 22000 iterations, discarding the first 2000 as the burn-in period and keeping every 20th draw. All prior distributions are analogous to the ones discussed in the first application, where we have specified noninformative prior distributions, with the exception of the nonlinear term, which is not used in this application. For this reason, we do not repeat this information here. \citet{guggisberg2017} discusses prior choices for this multiple-output problem, where the author show how one can define a prior belief that the response variable has spherical Tukey depth contours, which is the case for this application.

Given the posterior estimates, we are able to obtain quantile contours for a combination of covariates. These regions enable us to measure the effect of the covariates in different directions of the conditional distribution of the scores. Here we show these estimated effects for type of school and sex, but the same plots for all variables are available in the supplementary material. We also show an algorithm that one can use to produce these plots.

\begin{figure}[!ht]\centering
\includegraphics[width=0.3\textwidth]{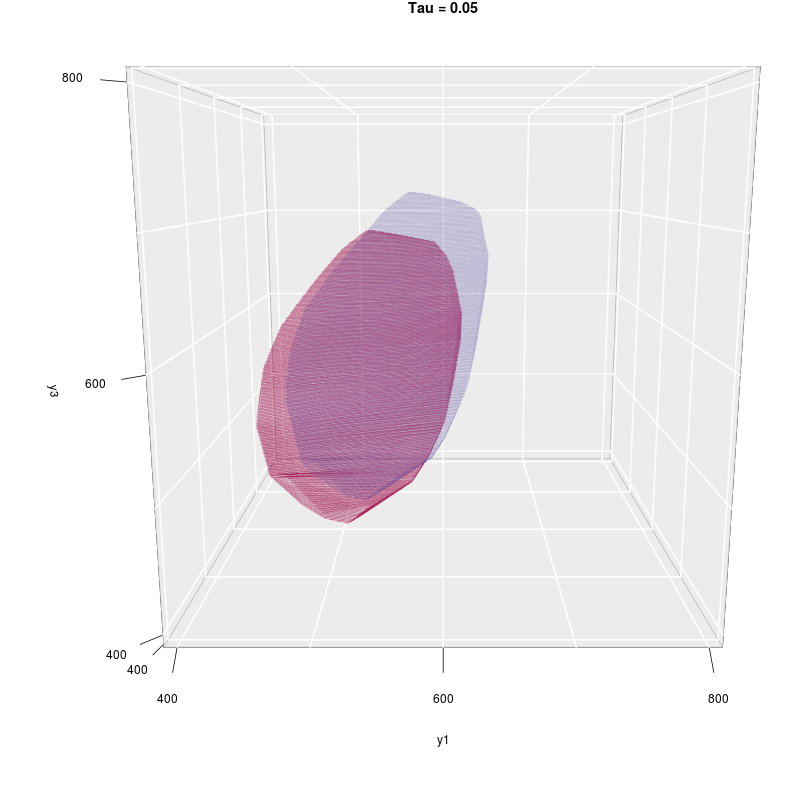}
\includegraphics[width=0.3\textwidth]{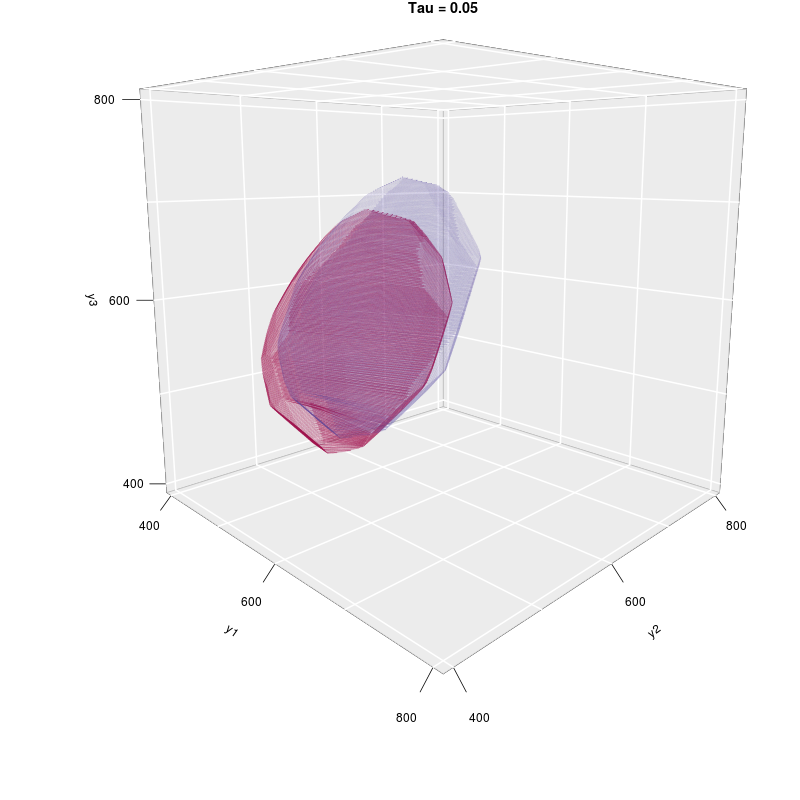}
\includegraphics[width=0.3\textwidth]{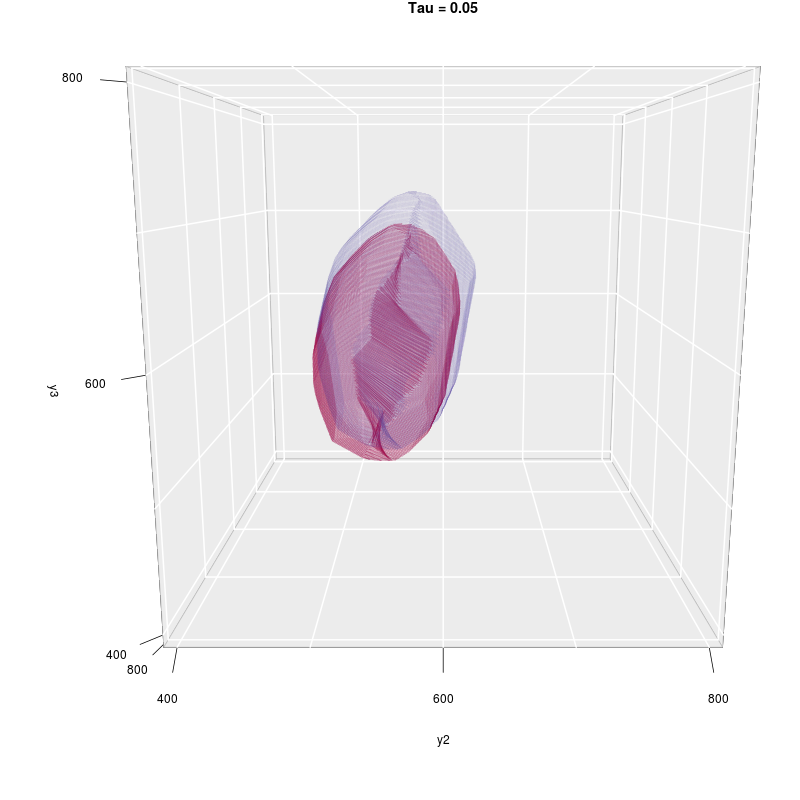}
\includegraphics[width=0.3\textwidth]{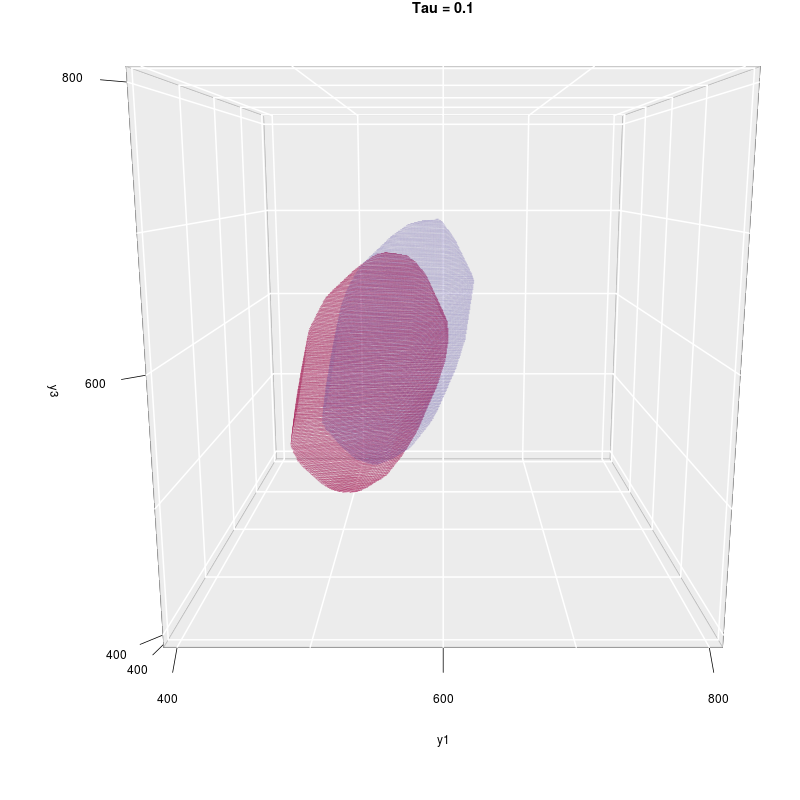}
\includegraphics[width=0.3\textwidth]{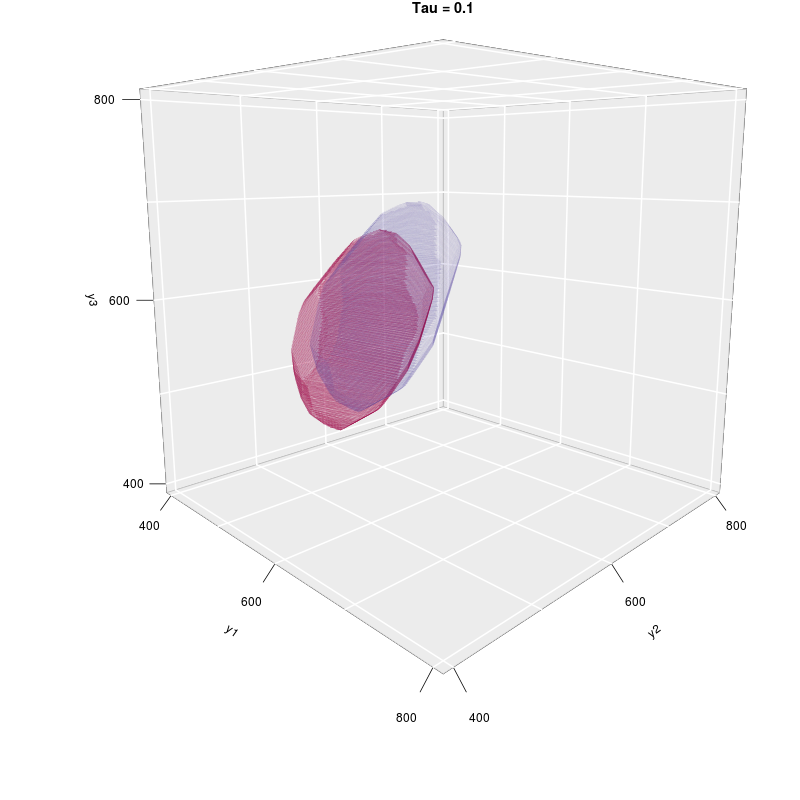}
\includegraphics[width=0.3\textwidth]{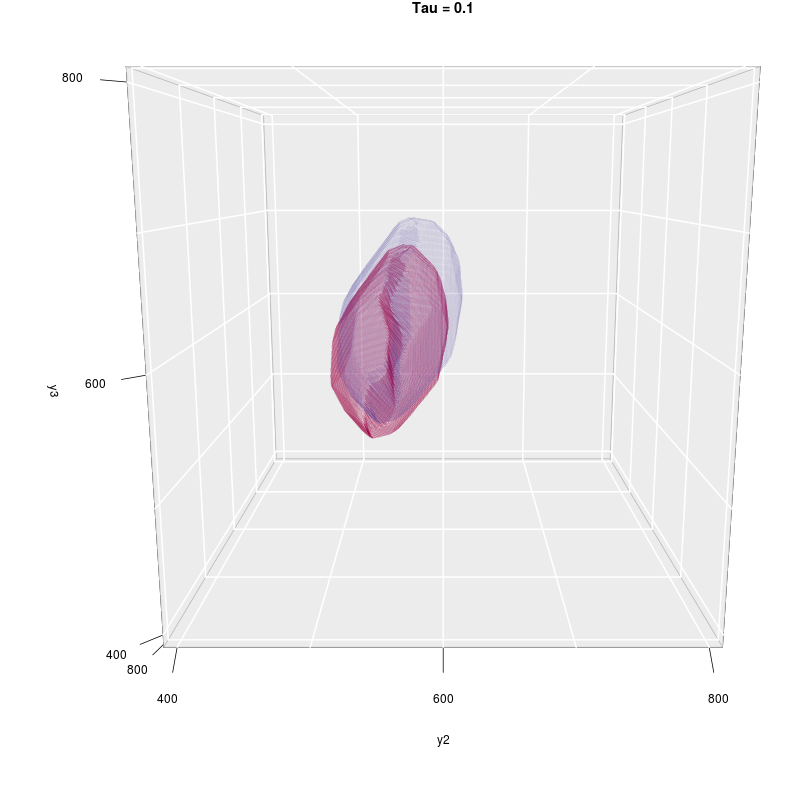}
\includegraphics[width=0.3\textwidth]{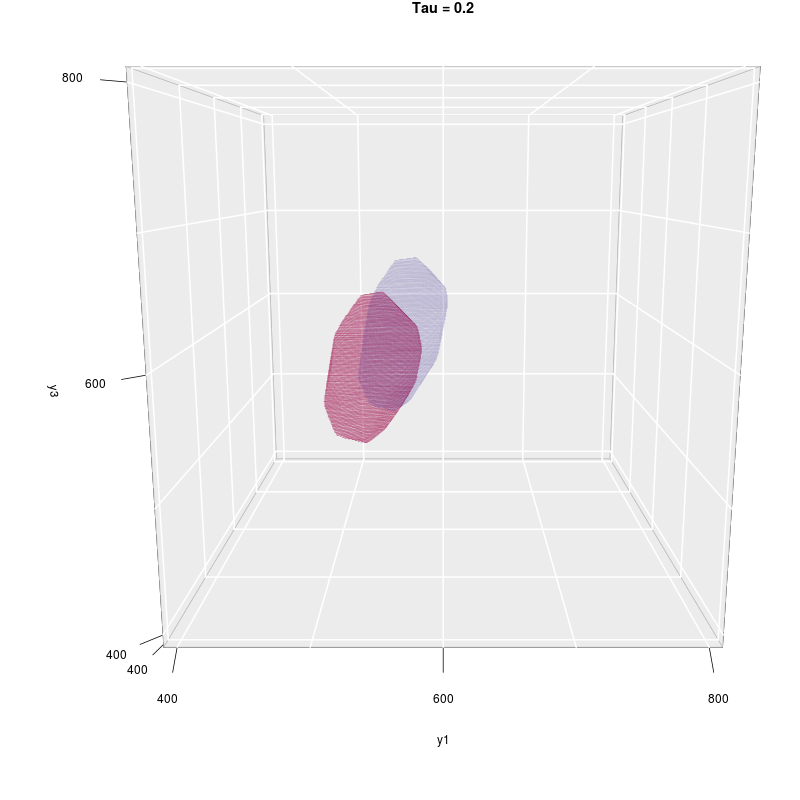}
\includegraphics[width=0.3\textwidth]{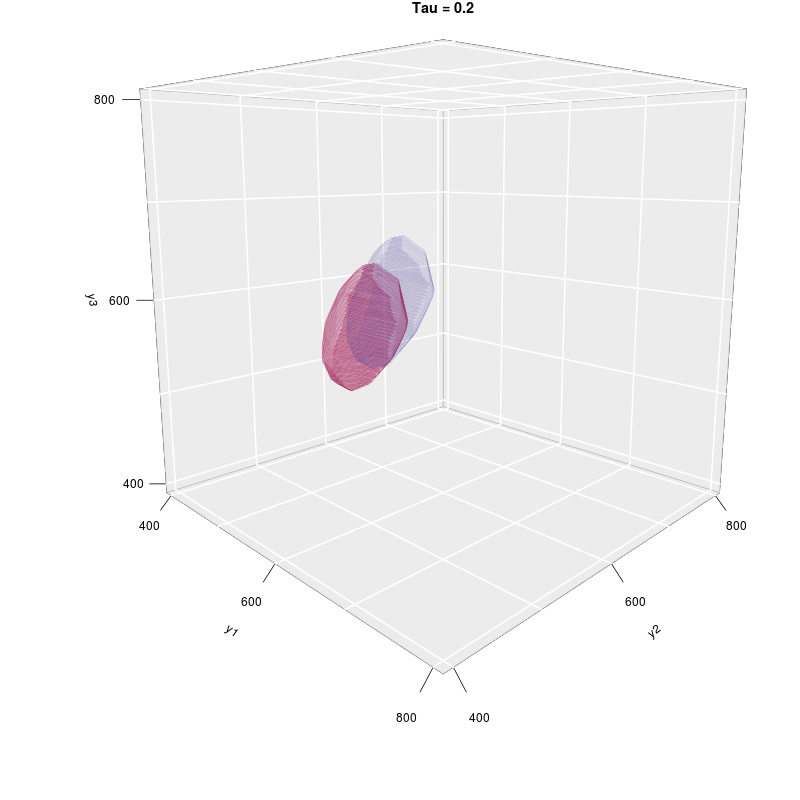}
\includegraphics[width=0.3\textwidth]{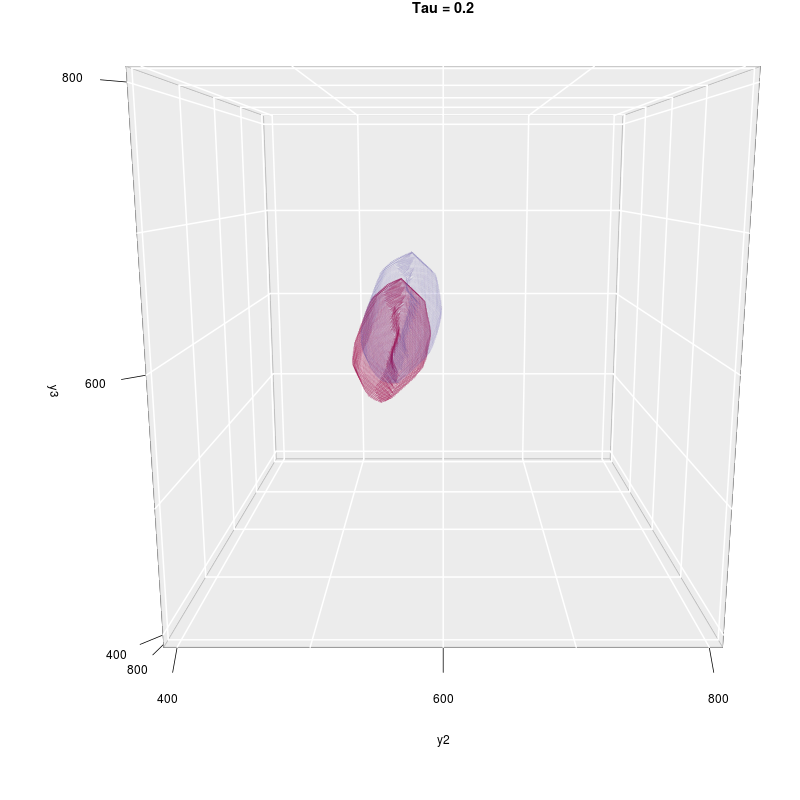}
\caption{\label{figure_effect_3d_school} Quantile contours for a comparison between private (blue) and public (red) for different values of $\tau$. In each row, the left plot shows the quantile contour for colatitude angle $\phi = 0^\circ$, the middle plot $\phi = 45^\circ$ and the right plot $\phi = 90^\circ$.}
\end{figure}

Regarding the results, one can see how scores for private schools are higher for most directions, in comparison with public schools, which is depicted in Figure~\ref{figure_effect_3d_school}. This difference explains, in part, why there was a necessity for creation of quotas on universities, in order to cope with these social inequalities \citep[see e.g.][]{mccowan:07}. This difference can be seen specially in the mathematics and natural sciences scores, for all quantiles. But this disparity is less pronounced for the human sciences scores for smaller quantiles and those students with lower scores. This indicates that between those students with a lower score in this discipline one cannot distinguish between public and private schools. This is another testament for this approach considering this multiple-output response variable, where one is able to view these multifaceted conclusions. 

\begin{figure}[!ht] \centering
\includegraphics[width=0.3\textwidth]{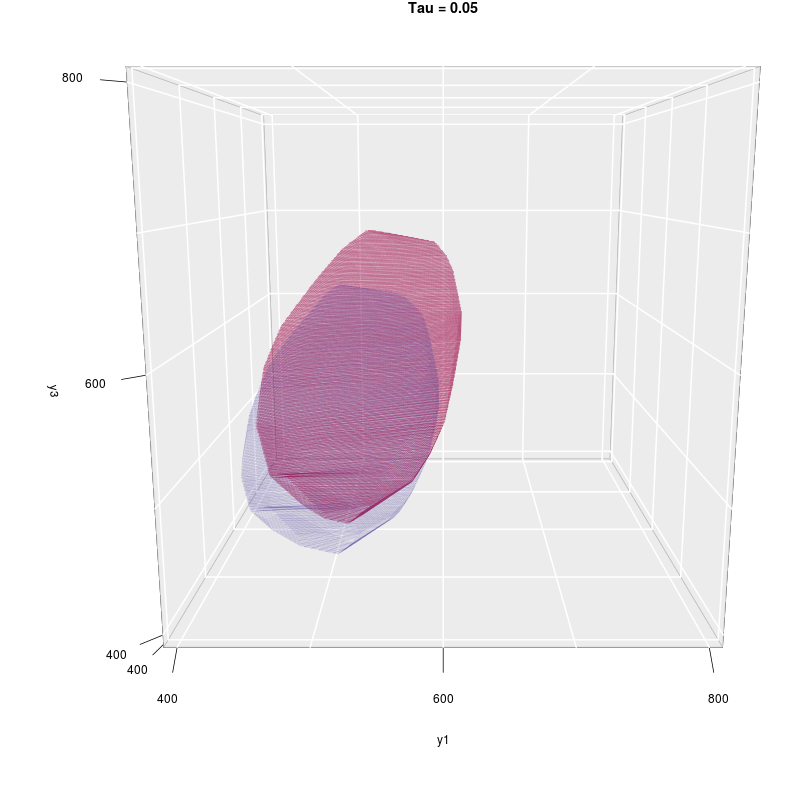}
\includegraphics[width=0.3\textwidth]{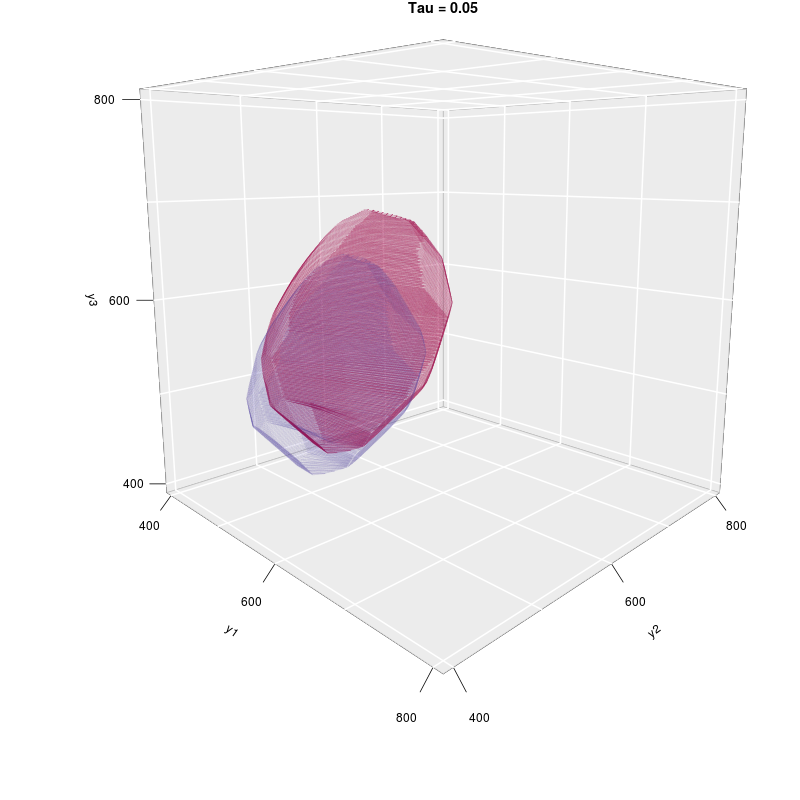}
\includegraphics[width=0.3\textwidth]{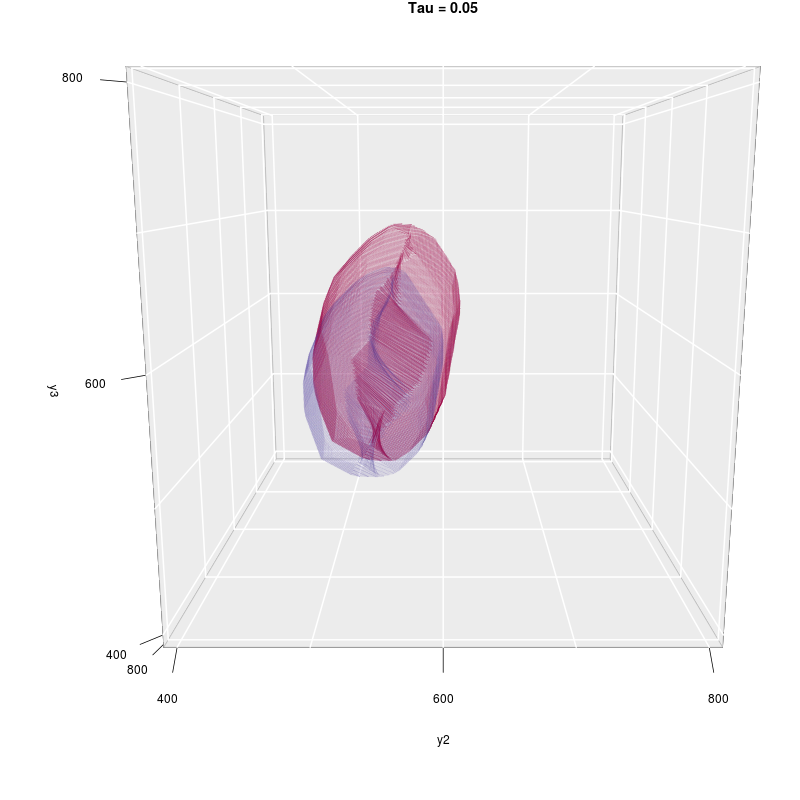}
\includegraphics[width=0.3\textwidth]{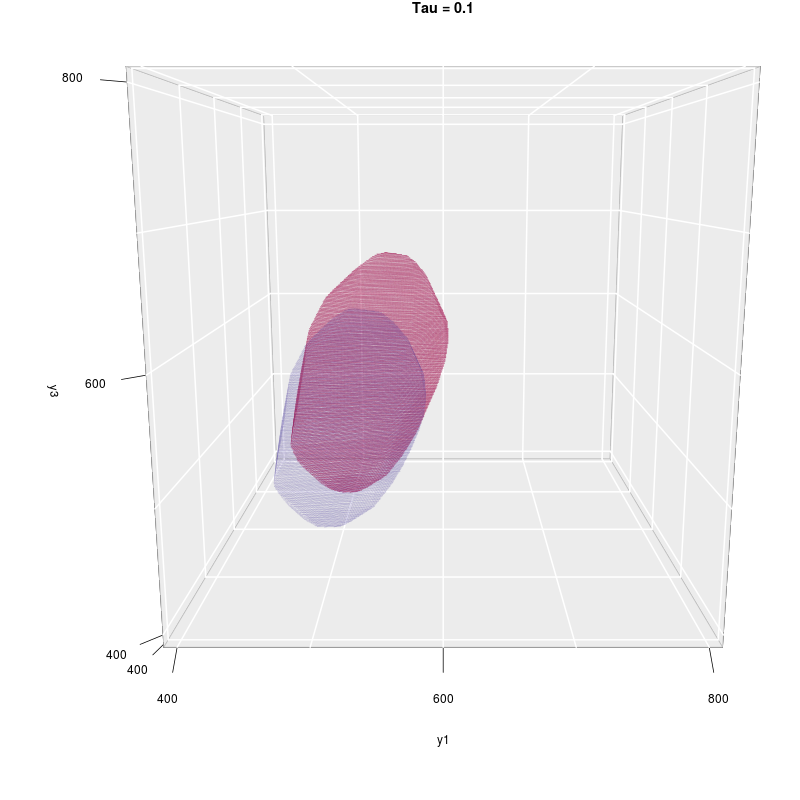}
\includegraphics[width=0.3\textwidth]{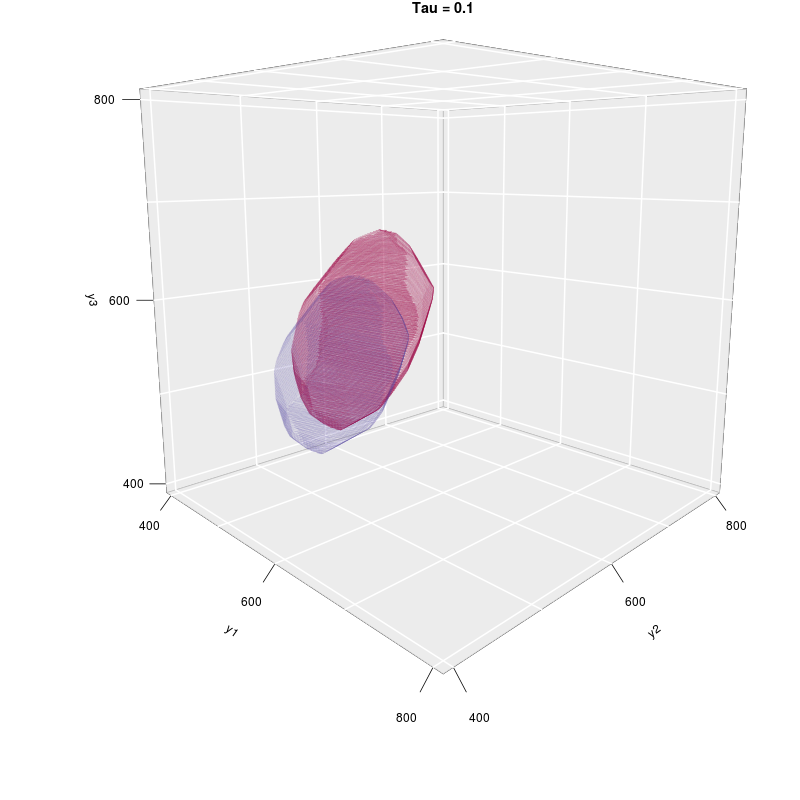}
\includegraphics[width=0.3\textwidth]{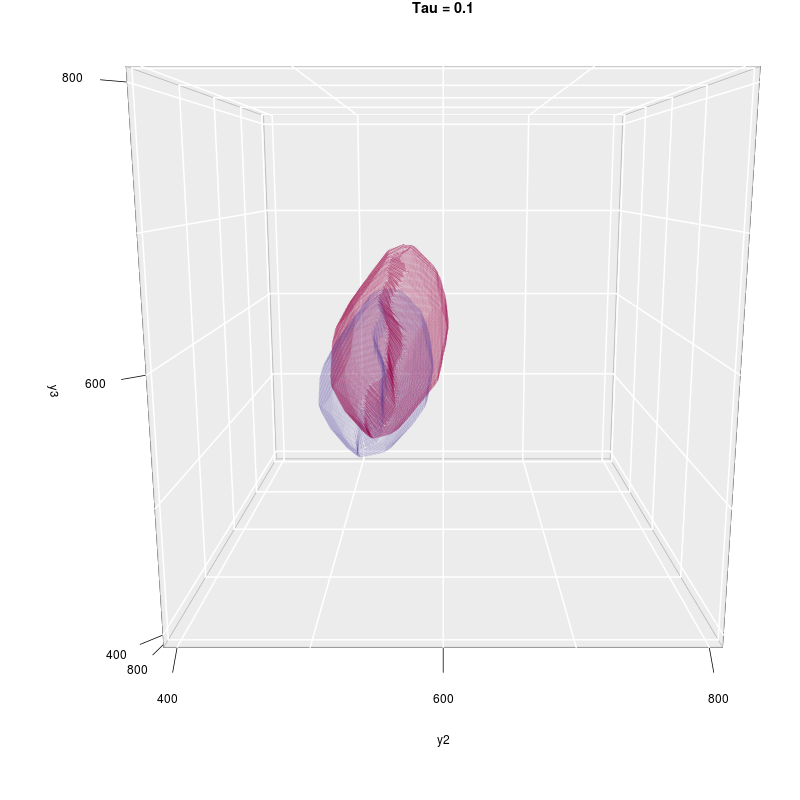}
\includegraphics[width=0.3\textwidth]{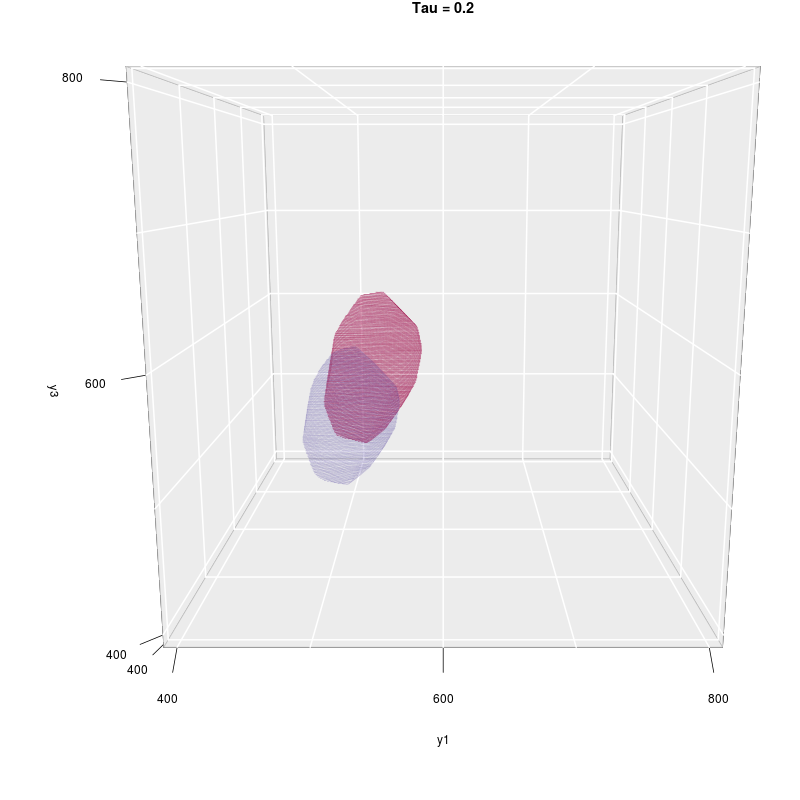}
\includegraphics[width=0.3\textwidth]{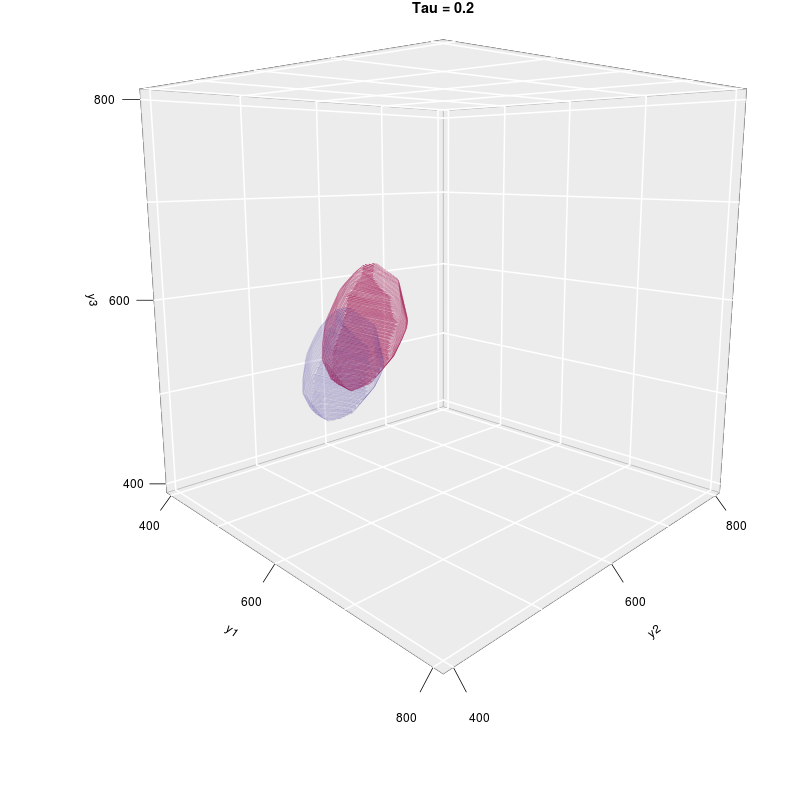}
\includegraphics[width=0.3\textwidth]{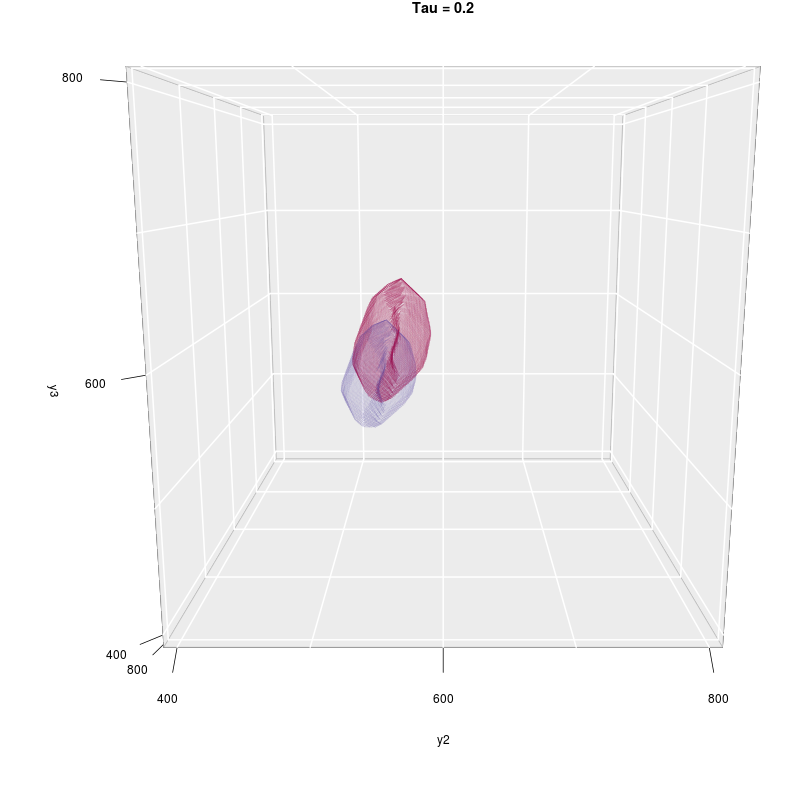}
\caption{\label{figure_effect_3D_gender} Quantile contours for a comparison between women (blue) and men (red) for different values of $\tau$. In each row, the left plot shows the quantile contour for colatitude angle $\phi = 0^\circ$, the middle plot $\phi = 45^\circ$ and the right plot $\phi = 90^\circ$. }
\end{figure}

Moreover, similar remarks can be made about the differences between men and women, which is shown in Figure~\ref{figure_effect_3D_gender}. Particularly, there is a great difference in the scores in mathematics between these two groups in all quantiles presented here. In addition, in the Supplementary Material one can see other interesting results, as we do not show the comparison for all covariates here for the sake of brevity. For instance, the education of the mother, in this case of the dummy variable for those who have a university degree, has a bigger impact on the student scores than the father education, with a similar coding. Also, variables such as income levels and race, which usually are important to describe social differences in Brazil, do not seem to impact these scores, when all the other variables are taken into account.

Finally, this application presents some interesting conclusions when one considers this multiple-output approach. Specifically for the multivariate distribution of the score in these three different disciplines, as presented in Figure~\ref{3D_data}. With this directional approach, one is able to illustrate the variation of its conditional distribution, identifying for which disciplines some differences can be considered more evident. While we made an effort to provide a more complete visualization of the quantile contours in this example, we emphasize that a better way of examine the effects of each variable is by having an interactive plot. With this kind of feature, one has the possibility of checking with more details the conclusions we briefly discuss here.

\section{Conclusion and discussion}
\label{finalRemarks}

With multivariate data comes the need to have methods that explore its dependences between variables in a reasonable way. In a regression setting, when the response variable has more than one dimension, it becomes harder to study its conditional distribution. Directional quantile regression models for multiple-output response variables is one option when one is interested in studying the conditional distribution of the response variables beyond central measurements, such as the conditional mean. This approach is directly connected to the notion of Tukey depth, which gives an idea of ordering for multidimensional random variables. This important result of this shared connection is specially exciting, given the usually complex nature of the algorithms proposed to calculate the Tukey depth. 

Here in this article we add the possibility of one taking a more flexible approach to defining how the predictor variables affect this multivariate response. This is possible considering structured additive predictors in the modeling process. This idea is accompanied by a Gaussian process regression adjustment, which guarantees noncrossing quantiles. Both ideas considered here were proposed in the literature for the univariate case, for which we made the necessary adjustments in the multivariate case. In an illustration, we showed how it might be important to take into account nonlinear functions to model certain types of predictor variables, such as age, due to the interest in estimation for several directions. Besides that, in this example we were able to identify how different combination of predictor variables might lead to very distinct comparison for the quantile contours. We also illustrated conditional quantile contours when the response variable has 3 dimensions, in the case of score exams in Brazil. 

Moreover, the ready access to information such as quantile contours shows that this method could be able to identify observations which lie more distant to others even in higher dimensions, when visualization becomes an issue, for instance. What is even more interesting is that this possible assignment is conditional on the predictor variables. This kind of check can be performed for every combination of predictor variables given a set of quantiles and respective quantile contours. The only problem when dealing with more dimensions is the necessity of choosing the directions $u$ for which the model is estimated. An option for this complication was proposed by \citet{carlier2016}, where the authors define a conditional vector quantile function, aimed for this multivariate outcome. This is done without the need to define a fixed set of directions and this approach could be a rich material for research in the future. 

Finally, with the advent of more models to deal with multivariate data comes the responsibility of visualizing and reporting the results. As we tried to discuss throughout the article, this is still a difficulty for more than 3 dimensions. We believe that this could be a focus for research as well and we are currently working to propose ideas in this direction.

\end{document}